\documentclass[sigconf]{acmart}

\def\Section {\S}

\AtBeginDocument{%
  \providecommand\BibTeX{{%
    \normalfont B\kern-0.5em{\scshape i\kern-0.25em b}\kern-0.8em\TeX}}}

\usepackage{xfrac}    
\usepackage{geometry}
\usepackage{amsmath}
\usepackage{siunitx}
\DeclareSIUnit\E{E}
\DeclareSIUnit\c{c}

\copyrightyear{2023}
\acmYear{2023}
\setcopyright{acmlicensed}\acmConference[e-Energy '23]{The 14th ACM International Conference on Future Energy Systems}{June 20--23,
2023}{Orlando, FL, USA}
\acmBooktitle{The 14th ACM International Conference on Future Energy
Systems (e-Energy '23), June 20--23, 2023, Orlando, FL, USA}
\acmPrice{15.00}
\acmDOI{10.1145/3575813.3595191}
\acmISBN{979-8-4007-0032-3/23/06}

\begin{document}

\title[Jointly Managing Electrical and Thermal Energy]{Jointly Managing Electrical and Thermal Energy in Solar- and Battery-powered Computer Systems}

\author{Noman Bashir, Yasra Chandio, David Irwin, Fatima Anwar, Jeremy Gummeson, Prashant Shenoy}
\affiliation{%
  \institution{University of Massachusetts Amherst}
  \streetaddress{}
  \city{}
  \country{}}

\renewcommand{\shortauthors}{Bashir et al.}

\begin{abstract}
Environmentally-powered computer systems operate on renewable energy harvested from their environment, such as solar or wind, and stored in batteries.  While harvesting environmental energy has long been necessary for small-scale embedded systems without access to external power sources, it is also increasingly important in designing sustainable larger-scale systems for edge applications. For sustained operations, such systems must consider not only the electrical energy but also the thermal energy available in the environment in their design and operation.  Unfortunately, prior work generally ignores the impact of thermal effects, and instead implicitly assumes ideal temperatures.  To address the problem, we develop a thermodynamic model that captures the interplay of electrical and thermal energy in environmentally-powered computer systems. The model captures the effect of environmental conditions, the system's physical properties, and workload scheduling on performance. In evaluating our model, we distill the thermal effects that impact these systems using a small-scale prototype and a programmable incubator. We then leverage our model to show how considering these thermal effects in designing and operating environmentally-powered computer systems of varying scales can improve their energy-efficiency, performance, and availability.
\end{abstract}

\ccsdesc[500]{Hardware~Thermal issues}
\ccsdesc[500]{Computer systems organization~Embedded and cyber-physical systems}
\ccsdesc[300]{General and reference~Design}
\ccsdesc[300]{General and reference~Performance}
\ccsdesc[300]{General and reference~Reliability}

\keywords{Environmentally-powered computer systems, thermal effects, energy-efficiency, performance, batteries.}

\maketitle

\section{Introduction}
\label{sec:introduction}
Environmentally-powered computer systems operate on renewable energy harvested from their environment, such as solar or wind, and stored in batteries.  
While harvesting environmental energy has long been necessary for small-scale embedded systems without external power sources~\cite{signpost,farmbeats}, it is also increasingly important in designing sustainable larger-scale systems for zero-carbon and edge applications.  For example, there has been a recent emphasis on designing zero-carbon edge-cloud infrastructure powered by renewable energy to mitigate climate change~\cite{hotnets21,socc21}.  

Since these systems' power is intermittent and limited by their environment, they must carefully regulate their energy usage over time to match their supply. 
While there has been substantial prior work on designing environmentally-powered systems that dynamically adapt their energy usage to enable perpetual operation, most of this work focuses on small-scale energy-harvesting sensor systems~\cite{perpetual,ganesan,cloudy}, which have little computing capacity, low computation density, and primarily focus on data sensing. 
However, the recent emergence of low-power and energy-efficient AI accelerators, such as NVIDIA's Jetson Nano~\cite{nano}, combined with the advances in solar and battery technologies is changing how these systems are designed and operated. 
The future environmentally-powered computer systems at the edge are going to be much more powerful running compute-intensive tasks such as AI inference and computer vision tasks. 
These workloads will originate from modern applications -- including precision agriculture, smart traffic monitoring and control, beehive/bird/animal population monitoring, live language translation, and others -- deployed in outdoor environments. 
These changes to system sizes, and the workloads that run on them, are increasing the importance of thermal effects, which have not been addressed in prior work. In general, environmentally-powered computer systems may be deployed in many different climates that subject them to a wide range of ambient temperatures, which can affect their operation in numerous and significant ways.

\begin{figure*}[t]
    \centering
    \begin{tabular}{ccc}
    \includegraphics[width=0.22\textwidth]{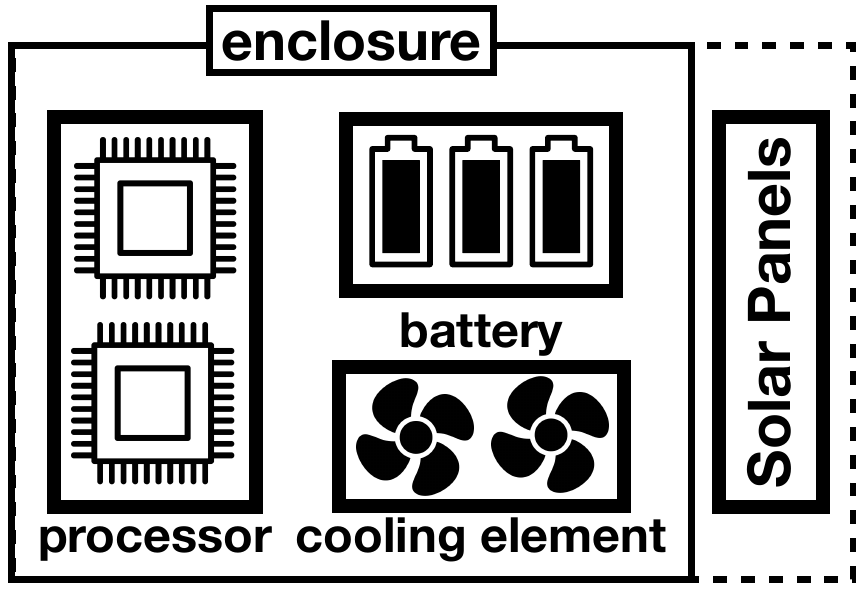} &
    \includegraphics[width=0.372\textwidth]{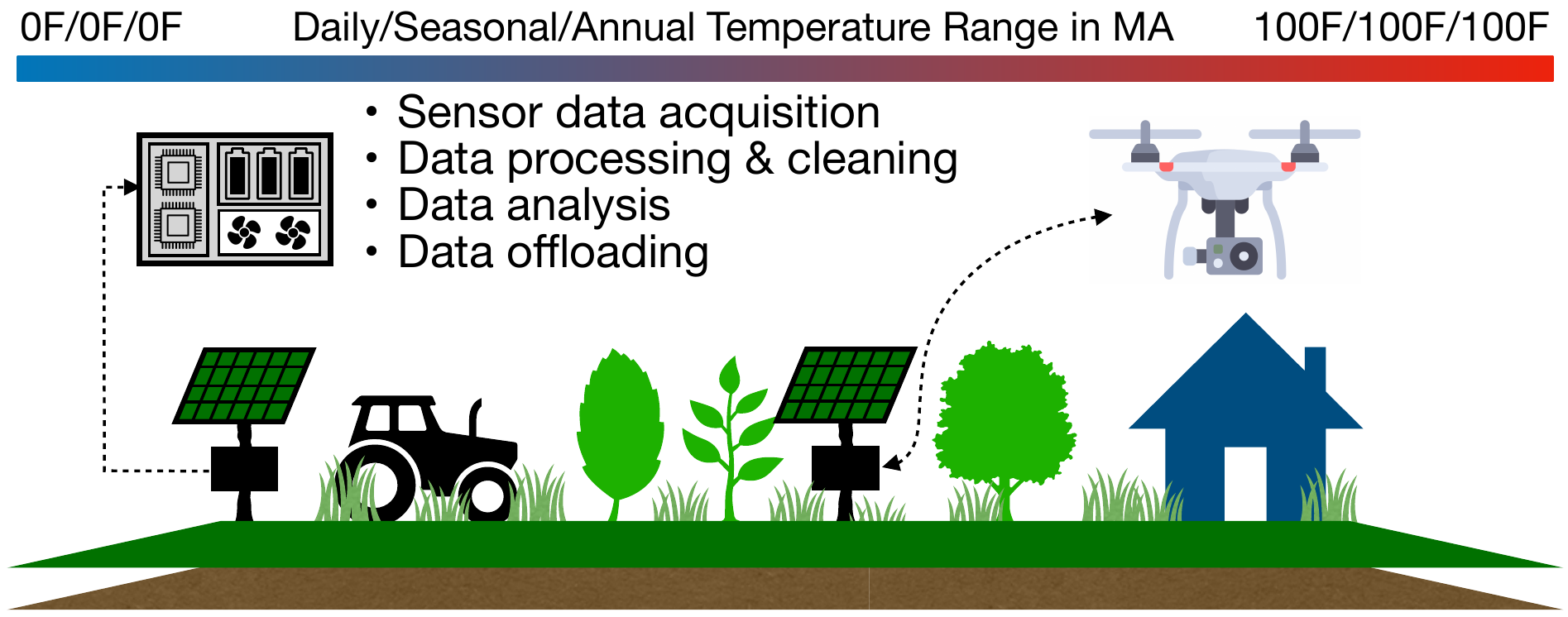} &
    \includegraphics[width=0.335\textwidth]{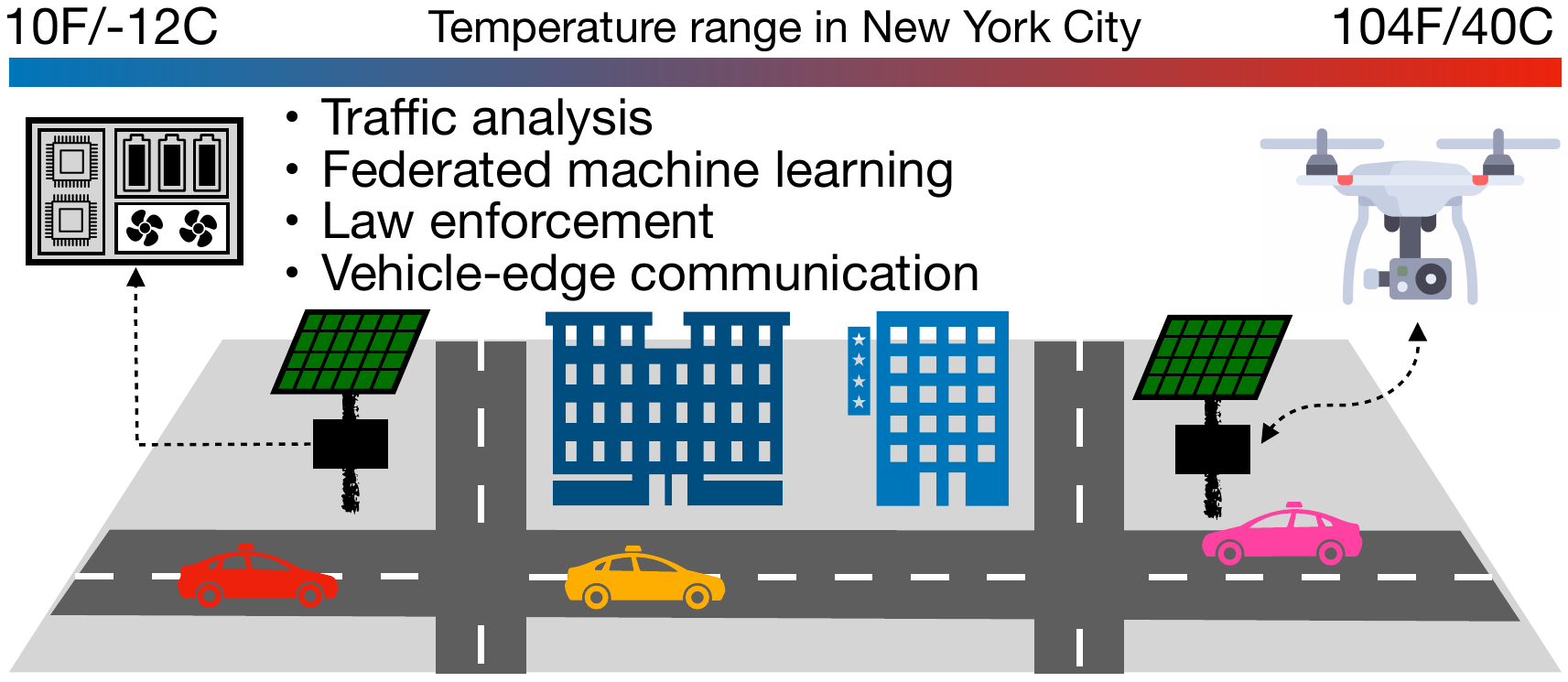} \\
    (a) System Components &
    (b) Precision Agriculture &
    (c) Smart Cities \\
    \end{tabular}
    \vspace{-0.5cm}
    \caption{\emph{Environmentally-powered computer systems consist of processors powered by solar and batteries and include a cooling element (a). Common applications include small- to medium-scale embedded systems. e.g., for precision agriculture (b) and medium- to large-scale edge data centers (c). In both cases, systems may be exposed to highly variable temperatures.}}
    \label{fig:application_figs}
    \vspace{-0.6cm}
\end{figure*}

In particular, changes in temperature alter the energy-efficiency of processors, batteries, and cooling elements in significant, but different, ways. For example, battery charging and discharging becomes much less energy-efficient as temperature decreases, and may shutdown if the temperature decreases or increases too much.  
Prior work has generally ignored these effects, and often implicitly assumes an ideal temperature range, e.g., 20-25$^\circ$C, even though most locations do not experience ideal temperatures year-round.  
The lack of consideration of thermal effects is one reason that reported uptimes for environmentally-powered systems, such as FarmBeats are often low, e.g., $<$30 days~\cite{farmbeats3}.  Our key insight is that, to optimize performance, energy-efficiency, and availability, environmentally-powered computer systems must jointly consider and manage both the electrical and thermal energy in the environment as part of their design and operation.  While much prior work has examined adapting system operation to match variations in available energy, e.g., from solar or wind, in both small-scale sensing systems~\cite{perpetual,ganesan,cloudy} and large-scale cloud systems~\cite{,greencassandra,parasol,greenslot,sharma:asplos11}, it has not addressed the significant impact of temperature on the solar- and battery-powered system design and operation.

To address the problem, we enumerate, quantify, and model the numerous thermal effects that impact solar- and battery-powered computer systems.  While the temperature responses of individual components, e.g., processors, batteries, cooling elements, etc., are well-known, optimizing the performance, energy-efficiency, and availability of these systems requires understanding the relationships between these components and their environment. For example, at low temperatures, environmentally-powered systems can leverage some of the thermal energy generated by their processors to heat their battery, which can significantly increase the energy-efficiency of both.  Of course, these systems must also efficiently dissipate their heat at high temperatures to prevent processors and batteries from over-heating and becoming unavailable. 

To this end, we design a thermodynamic model for environmentally-powered systems by combining well-known physical models of heat transfer, batteries, processors, and cooling elements. 
Importantly, our model captures \emph{thermal feedback loops} between components that affect the system's operation, such as how scavenging a system's waste heat warms its battery, increasing its energy-efficiency, and enabling more computation. We empirically validate our model using a small-scale prototype and programmable incubator that precisely regulates temperature between -30$^\circ$C and 40$^\circ$C. We then leverage our thermal model to show how considering thermal effects in both designing and operating environmentally-powered computer systems can improve their performance, energy-efficiency, and availability. 

Specifically, our model and analysis quantifies the effect of a system's power draw, enclosure insulation, and ambient temperature on its energy-efficiency, i.e., computational work done using a fully charged battery.  We also highlight the tradeoff between an enclosure's heat transfer coefficient and its energy-efficiency: better insulation increases energy-efficiency when cold by more productively using waste heat, but decreases it when hot by requiring additional energy to power a cooling element that dissipates heat to prevent battery and processor over-heating.  Our work differs from prior work on optimizing the cooling infrastructure of data. centers powered by the electric grid, as that work mostly focuses on the efficient movement of heat from \emph{within} the facility to outside of it, and does not exhibit the feedback loop between computation and batteries present in environmentally-powered systems.

Our work demonstrates that managing and adapting to variable thermal energy is just as important as electrical energy in solar- and battery-powered computer systems, and that they are dependent on each other.  Currently, thermal management is mostly an after-thought for these systems with most implicitly designed for ideal-to-higher temperatures, often with little insulation that reduces the need for active cooling as temperatures rise, but wastes much of the heat these systems produce as temperatures drop.  There is currently little understanding, and no explicit modeling, of how temperature affects these systems. Our work is an important step towards better understanding how the temperature effects of individual components manifest at the system-level.  

Our hypothesis is that optimizing environmentally-powered computer systems requires jointly managing their electrical and thermal energy as part of their design and operation.  In evaluating our hypothesis, we make the following contributions.

\noindent {\bf Thermodynamic Model and Validation}.  We design a comprehensive thermodynamic model for an environmentally-powered computer system that consists of an energy source, e.g., solar panel, enclosure, batteries, and processors that are subjected to some ambient temperature.  The model accounts for the effect of heat and processor power on battery capacity, charging, and discharging, the heat emitted by the processor, and the energy consumed by a cooling element to dissipate heat. We validate our model by enumerating, isolating, and empirically quantifying the thermodynamic effects that impact the system's operation, and how they relate to each other.  Our empirical analysis demonstrates the impact of each effect on system operation under different ambient temperatures. 

\noindent {\bf System Design and Operation Use Cases}.  We present both a design and operation use case for our thermodynamic model.   In the design use case, we leverage our thermodynamic model to highlight the tradeoffs between system design parameters and user-specified performance objectives, e.g., for performance, energy-efficiency, and availability. In the operation use case, we demonstrate how a scheduler can leverage our thermodynamic model to operate the given design of an environmentally-powered computer system to optimize for a user-specified performance objective.

\noindent {\bf Implementation and Evaluation}. We implement a small-scale prototype and programmable incubator to empirically validate our model. We develop a model-driven simulator to enable long-term experimentation. We quantify the design tradeoffs and the operational space to show how our thermodynamic model can be leveraged to improve the system-level performance of two case study applications---a small-scale embedded system for precision agriculture and medium-scale federated learning at an edge datacenter. 

\section{Background}
\label{sec:background}
We summarize the thermodynamic effects exhibited by batteries, processors, and cooling elements, and how they alter the energy-efficiency of each.  We model these effects in the next section.

\vspace{-0.15cm}
\subsection{Environmentally-powered Systems}
Environmentally-powered computer systems operate on renewable energy harvested from their environment, such as solar or wind, and stored in batteries. 
Figure~\ref{fig:application_figs}(a) shows these systems' typical components,  including solar panels, processors, batteries, and a cooling element, such as a fan or pump. Figure~\ref{fig:application_figs}(b) $\&$ (c) show two example applications that leverage environmentally-powered systems. Precision agriculture applications deploy these systems to gather data from small-scale embedded devices that monitor environmental conditions, such as soil moisture, humidity, and temperature.  Similarly, there are a wide range of smart city applications, such as traffic monitoring, vehicle-to-edge communication, and crime detection, that analyze and process the data collected by sensors and cameras at medium- to large-scale edge data centers. 

\vspace{-0.15cm}
\subsection{Batteries} 
\vspace{-0.05cm}
The energy stored by batteries is related to their temperature and discharging/charging current.  We discuss these relationships below. 

\begin{figure}[t]
    \centering
    \begin{tabular}{c}
    \includegraphics[width=0.75\linewidth]{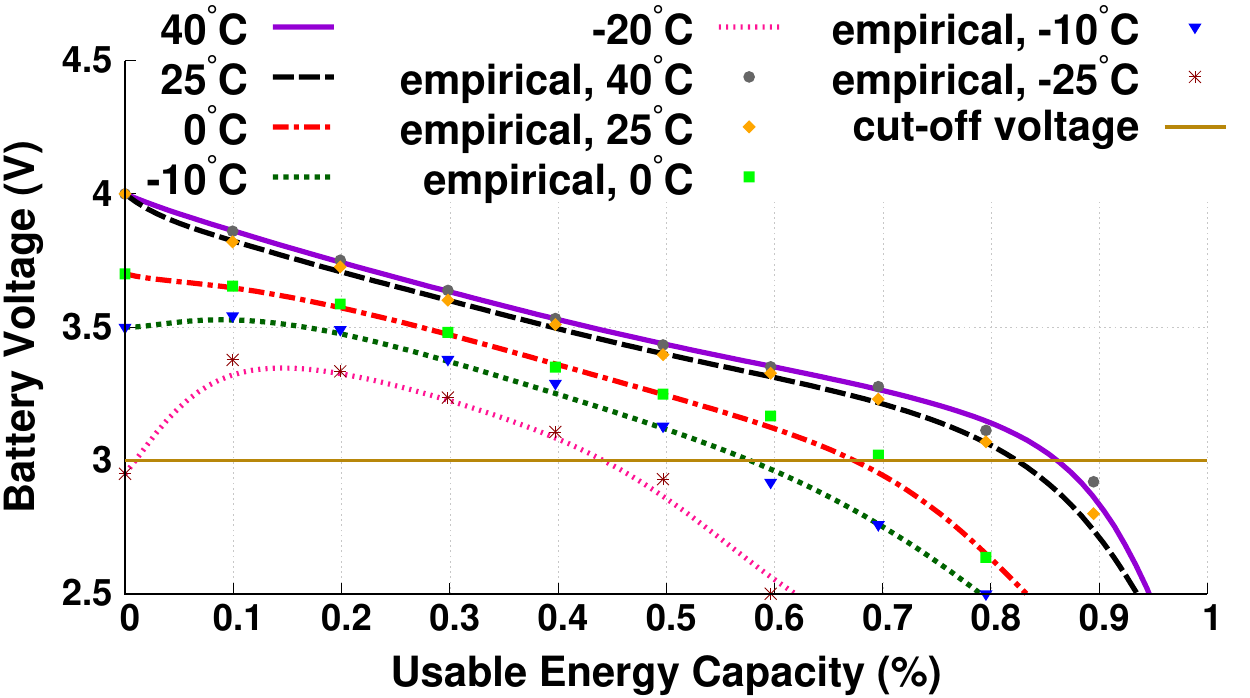} \vspace{-0.05cm} \\
    (a) Effect of Temperature \\ 
    \vspace{0.1cm}
    \includegraphics[width=0.75\linewidth]{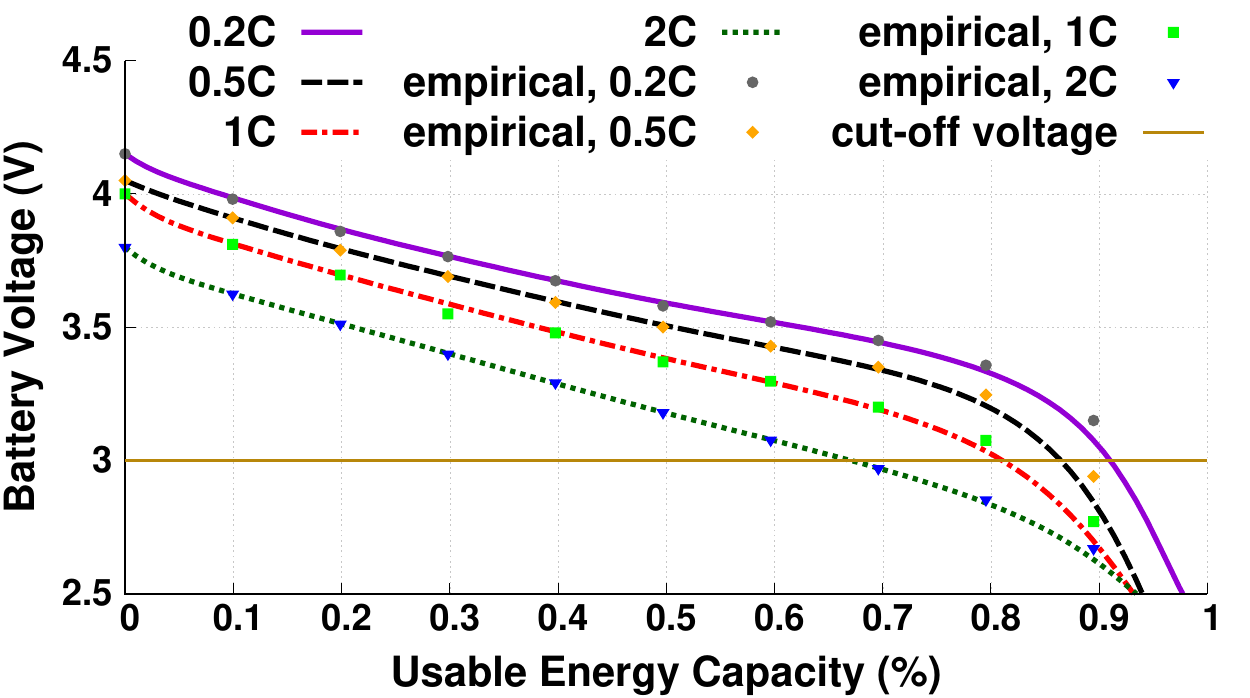}\vspace{-0.1cm}\\
    (b) Effect of Discharge Current \\
    \end{tabular}
    \vspace{-0.4cm}
    \caption{\emph{A system's usable battery capacity varies with both temperature (a) and discharge current (b).}}
    \vspace{-0.8cm}
    \label{fig:discharge_characteristics}
\end{figure}

\noindent{\bf Temperature-Energy Effect.}
The usable energy capacity of lithium batteries decreases with temperature.  Figure~\ref{fig:discharge_characteristics}(a) shows curves from our prototype's battery datasheet, where the points represent experiments we run to empirically validate the datasheet using our programmable incubator.  
The graph shows that the battery's usable capacity, as a percentage of its charged capacity, drops by over 50\% at the 3V cut-off voltage when the temperature drops from 25$^\circ$C to -20$^\circ$C.  
This ``wasted'' energy is consumed as heat by the battery to catalyze its chemical reaction that produces electricity.  
We call this the \emph{temperature-energy} effect. In addition, discharging at low temperatures can damage batteries and reduce cycle-lifetime.  In general, lithium batteries should not be discharged below -20$^\circ$. Likewise, as temperatures increase, batteries' self-discharge rate also increases, which reduces the energy available for discharge, although not by as much as a decrease in temperature.  Lithium batteries generally cannot be discharged at $>$60$^\circ$, and as at that temperature their available capacity drops to $0$. 

\begin{figure*}[t]
    \centering
    \begin{tabular}{ccc}
    \includegraphics[width=0.3\textwidth]{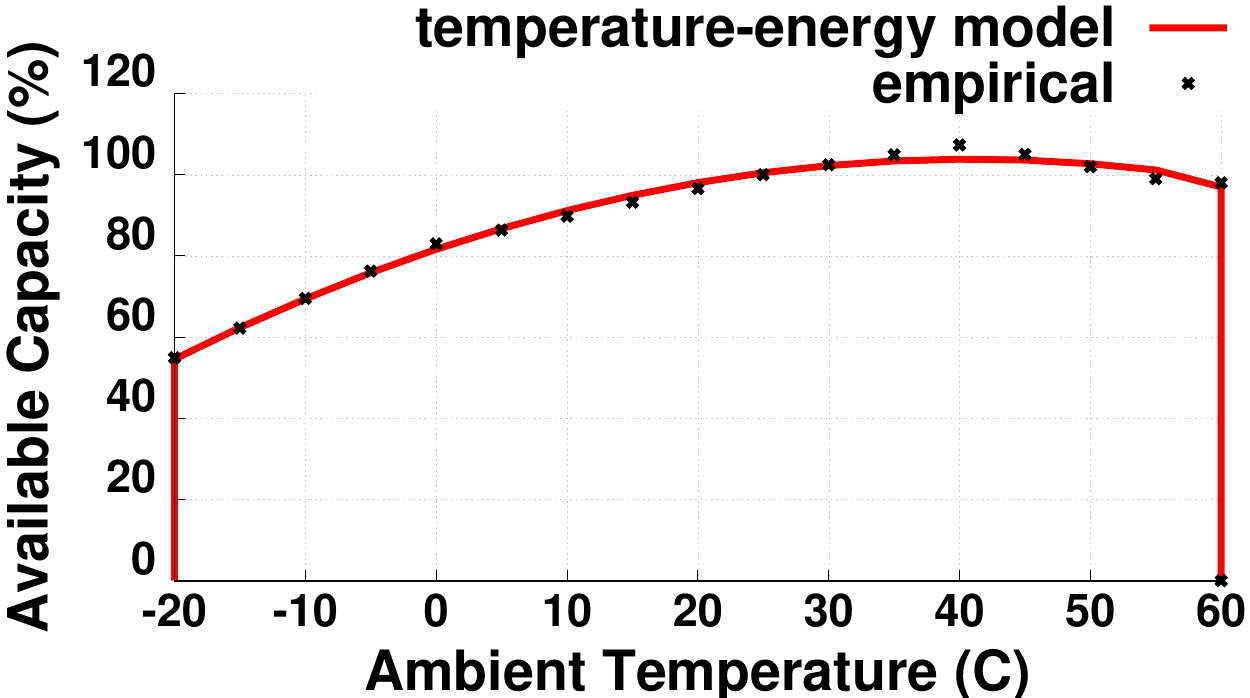} &
    \includegraphics[width=0.3\textwidth]{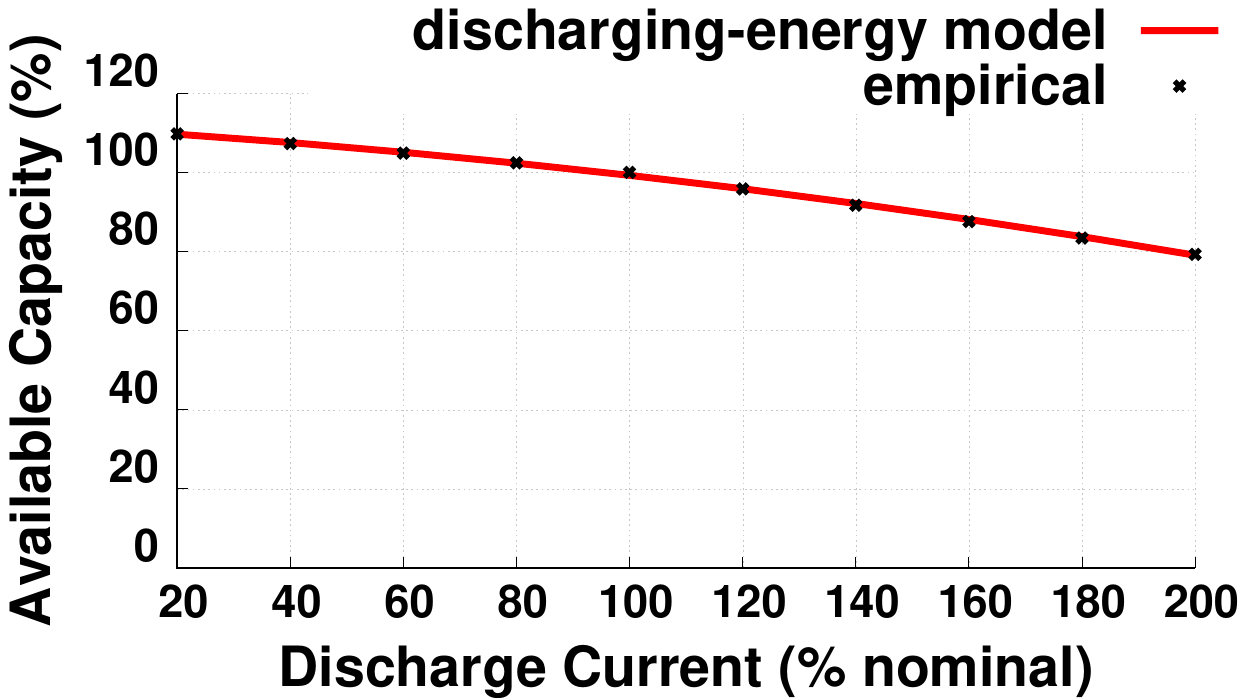} &
    \includegraphics[width=0.3\textwidth]{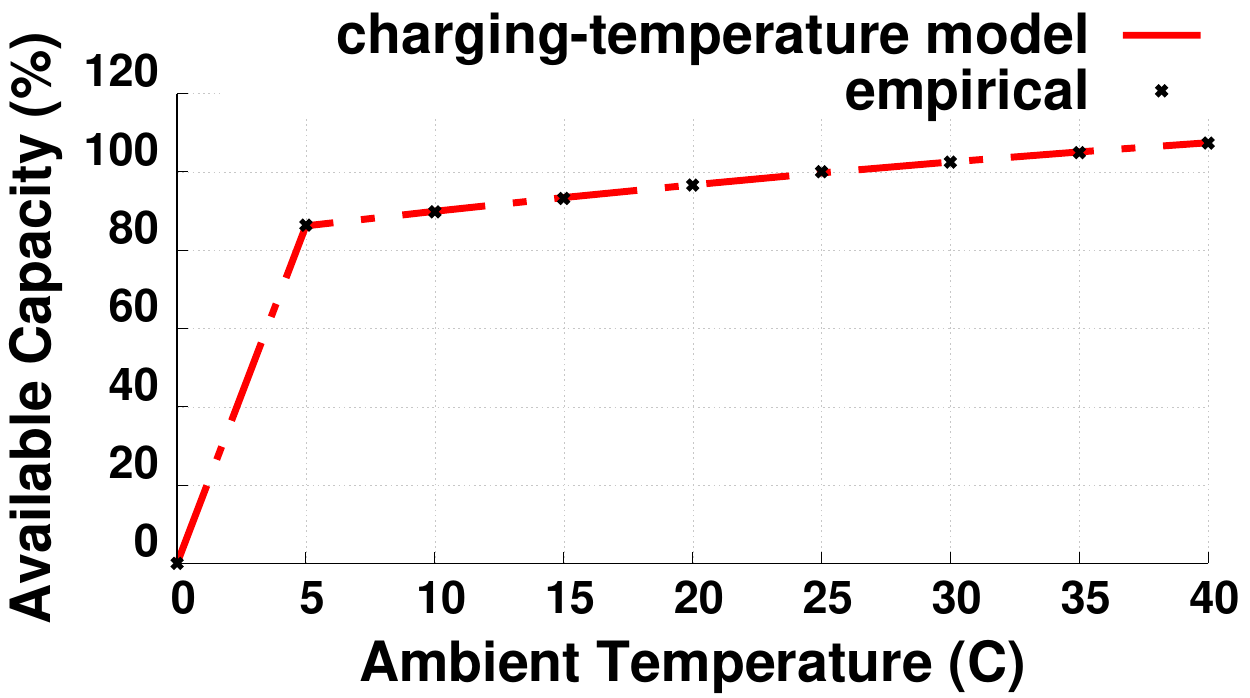}\\
    (a) Temperature-Energy Effect&
    (b) Discharge-Energy Effect&
    (c) Temperature-Charging Effect\\
    \end{tabular}
    \vspace{-0.35cm}
    \caption{\emph{Energy-efficiency as a function of temperature (a), discharge current (b), and available battery capacity when charging at various temperature (c). For (a) and (b), the black points are experimental data and the red curves represent our model.}}
    \label{fig:battery-effects}
    \vspace{-0.45cm}
\end{figure*}

Figure~\ref{fig:battery-effects}(a) plots the temperature-energy effect in our prototype with ambient temperature on the x-axis and available battery capacity on the y-axis.  The points represent experiments using our prototype, while the continuous line represents our model's prediction, which closely matches the data.   
In this case, we set the battery to 100\% capacity at 25$^\circ$C. As shown, the available capacity drops significantly as the temperature decreases, with only 80\% of the energy available at 0$^\circ$C and 50\% available near -20$^\circ$C. Temperatures between 0$^\circ$C and -20$^\circ$C are common over winter in much of the U.S., Europe, and other high latitude locations. As we show, the heat generated by processors can be leveraged to raise the internal enclosure temperature and extract more energy from batteries.  
The plot also shows how our prototype's charge controller automatically shuts down the battery once it reaches 60$^\circ$C for safety.  
While ambient temperatures generally do not reach 60$^\circ$C, they can reach this high within an insulated enclosure if processors generate heat faster than the system can dissipate it, i.e., via convection. 

\noindent{\bf Discharge-Energy Effect.}
The \emph{discharge-energy effect} refers to the decrease in available energy that occurs when discharging at higher rates. Figure~\ref{fig:discharge_characteristics}(b) shows curves from our prototype's battery datasheet, where the points represent experiments we have run for empirical validation.  The graph shows that, at the 3V cut-off voltage, the battery's usable energy decreases by 25\% when increasing the current from a C-rate of $0.2$ to $2$, where a C-rate of $N$ represents the discharge current required to fully discharge the battery in $1/N$ hours.  Due to this effect, if processors execute at 100\% utilization, they draw less energy from a battery than if they operate at lower utilization, since utilization is roughly linear with current draw.  Thus, the slower processing speed, the more energy they can extract from batteries, and the more overall computation they can perform.  

Figure~\ref{fig:battery-effects}(b) quantifies the discharge-energy effect where we plot the discharge current (which is linear with utilization) on the x-axis and the available energy capacity on the y-axis.  The points represent experiments with our prototype, while the continuous line represents our model's prediction, which closely matches the empirical data. Here, we normalize the experiment to some discharge current (equivalent to 50\% utilization), which we set at 100\%, and then set the available capacity at 100\% for that discharge current. 
This setting allows us to evaluate the discharge-energy effect at currents higher than 1C, which may be required by the system to server workload bursts. 
The experiment shows a linear relationship: as we slow down the system (by reducing utilization), we are able to extract more than nominal available energy, and as we speed up the system (by increasing its utilization), we draw less energy. In this case, only 80\% capacity is available when operating at 100\% utilization compared to 50\% utilization. The discharge-energy effect counteracts the temperature-energy effect:  running the processor at high utilization generates more heat, which can warm the battery and increase its available energy, but the increased discharge current reduces the available energy. Thus, determining the most energy-efficient operating point at any given time is non-trivial. 

\noindent{\bf Temperature-Charging Effect.} 
The \emph{temperature-charging effect} refers to the relationship between temperature and the rate at which batteries can charge.  While lithium batteries can safely discharge down to $-20^\circ$C, their maximum charging rate decreases with temperature and charging is not possible below $0^\circ$C.   Thus, \emph{low temperatures prevent storing energy and make using energy much less efficient}.   Lithium batteries can also easily overheat at high temperatures, since their chemical reaction generates additional heat that increases their internal temperature beyond the ambient temperature. Charge controllers generally prevent charging/discharging when the internal temperature rises too high (above $\sim$$60^\circ$C). Thus, \emph{high temperatures can prevent storing and using energy.}  Figure~\ref{fig:battery-effects}(c) shows how the charging capacity of the battery changes with temperature.  In this case, 100\% represents the maximum charging capacity at 25$^\circ$C. As shown, charging rate decreases rapidly from 5$^\circ$C to 0$^\circ$C, where it falls to 0\%.  At 5$^\circ$C capacity is $\sim$80\% and then increases roughly linearly with temperature.  As expected, higher temperatures enable the system to charge the battery at faster rates. As above, the points represent experiments using our prototype and line represents our model's prediction, which closely matches.

\vspace{-0.2cm}
\subsection{Processors}   
\vspace{-0.05cm}
Since processors do no mechanical work, their power is converted to heat, which must be dissipated to prevent them from overheating due to thermal runaway.  A system's energy-efficiency is a function of temperature if it leverages outside air (or water) for cooling, since lower ambient temperatures require using less additional energy to actively cool the processor. Such ``free cooling'' is often used by cloud data centers~\cite{free-cooling1}. Thus, unlike batteries, processors are \emph{more} energy-efficient at low temperatures, since they do not have to consume additional power to dissipate heat.  Also unlike batteries, computer systems (at a fixed frequency and voltage) are more energy-efficient, in joules per computation, at higher power, since they are generally not energy-proportional and a higher power (and utilization) amortizes their baseload power over more computation.
Of course, an ideal energy-proportional system has the same energy-efficiency at any utilization. The heat generated by processors can be recycled by environmentally-powered systems to optimize the efficiency of the battery based on the various effects above.  For example, scheduling workload at low temperatures can improve system performance by generating heat that improves battery efficiency.  We call this the \emph{scheduling-performance} effect.

\vspace{-0.2cm}
\subsection{Cooling System}  
\vspace{-0.05cm}
Fans (or pumps) are generally used to dissipate heat in computer systems.  The intensity with which a variable speed motor must rotate the blades of a fan (or pump) to maintain a certain temperature is a function of the heat dissipated by the processor, the conductivity of the system enclosure's insulation, and the external temperature.  Thus, there is an \emph{insulation-cooling effect} that impacts system enclosure design:  the thicker the enclosure's insulation, the better its performance in low temperatures, but the more the motor must run in high temperatures, and vice versa.   In addition, as we discuss, fan (and pump) energy usage is a cubic function of the amount of air (or water) it moves, and thus the heat it dissipates.

\begin{figure}[t]
    \includegraphics[width=0.75\linewidth]{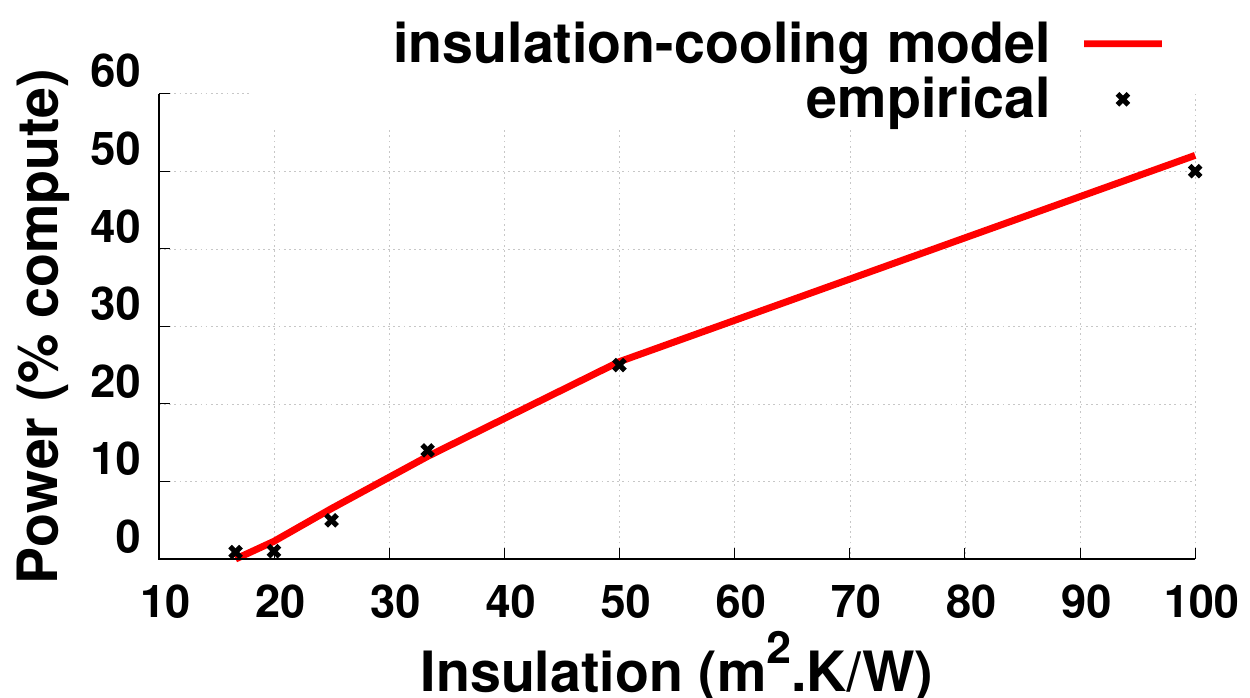}
    \vspace{-0.34cm}
    \caption{\emph{Fan power as a function of insulation. Black points are experimental data and the red curve represents our model.}}
    \label{fig:insulation-fan-effect}
    \vspace{-0.6cm}
\end{figure}

The \emph{insulation-cooling effect} captures the tradeoff between having thicker insulation to retain heat during low temperatures at the cost of consuming more energy via a fan (or pump) to dissipate heat at high temperatures. Figure~\ref{fig:insulation-fan-effect} quantifies the insulation-cooling effect for our prototype at 25$^\circ$C for different levels of insulation on the x-axis.  The y-axis shows the power required by the fan (or pump) motor to maintain 25$^\circ$C as the enclosure's insulation increases when operating the system at 50\% utilization.  As expected, when the insulation is thin, there is almost no need for cooling, and it consumes little power.  However, as we increase the insulation's thickness, the enclosure retains more heat, which requires consuming more power to dissipate that heat by active cooling. Of course, the energy used by the cooling element is energy that does not go towards productive computation. One option at these higher insulation levels, instead of running the cooling element, is to operate at a lower utilization to generate less heat, which reduces the need to consume energy by the fan to dissipate heat.  Thus, as with the discharge-energy effect, operating slower, at a lower utilization, enables more energy to go towards productive computation.  
That is, active heat dissipation using the cooling element enables environmentally-powered systems to operate at higher workload intensities than they otherwise could, but at the cost of lower energy-efficiency.  
Also, as above, the points in the graph represent experimental data from our prototype, while the line represents our model's prediction, which closely matches. 

\section{Thermodynamic Model}
\label{sec:thermodynamic_model}

\begin{figure}[t]
    \includegraphics[width=0.95\linewidth]{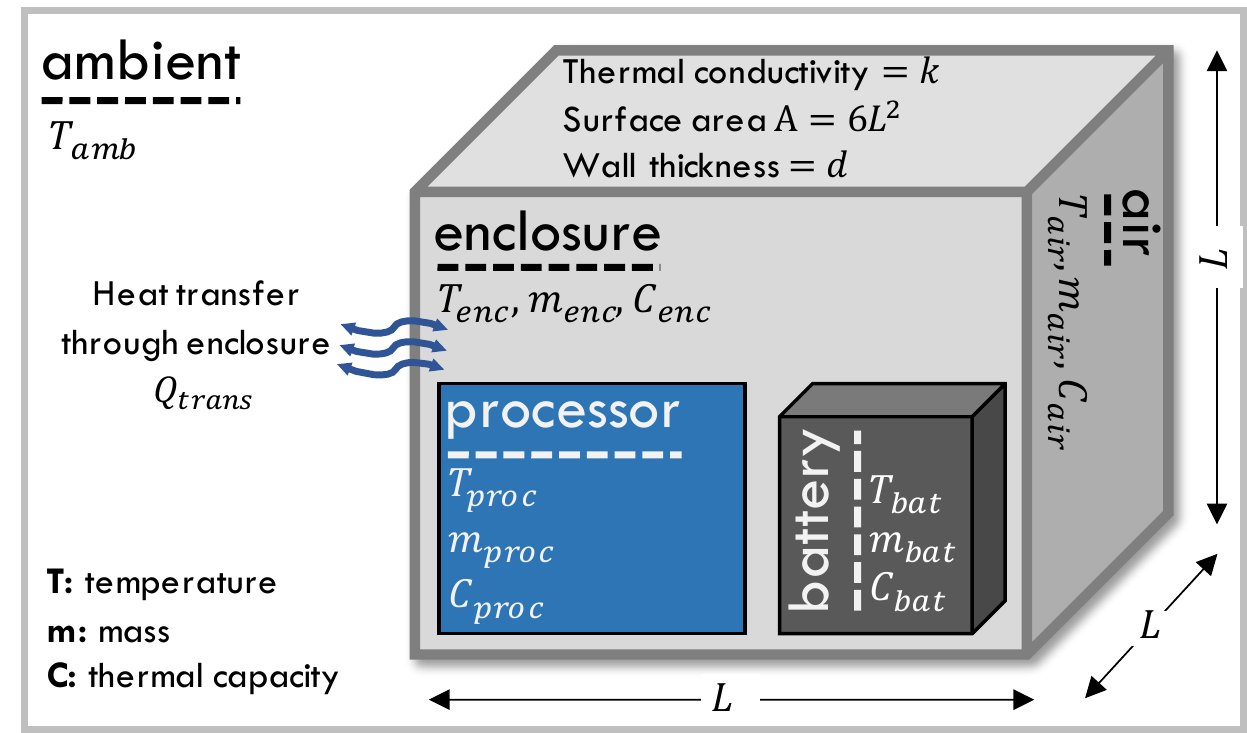}
    \vspace{-0.3cm}
    \caption{\emph{A simple and general thermodynamic model of an environmentally-powered computer system.}}
    \label{fig:box}
    \vspace{-0.4cm}
\end{figure}

To better understand the effect of temperature on the operation of an environmentally-powered system, we develop a comprehensive physical thermodynamic model that estimates a system enclosure's change in temperature over some time interval $\Delta t$. Our contribution lies in leveraging basic thermodynamic relationships to develop an end-to-end model for predicting system-level performance; the basic relationships can be found in classic thermodynamic textbooks~\cite{thermo-book1, thermo-book2}. Figure~\ref{fig:box} illustrates our model and its key parameters. Table~\ref{tbl:notations} outlines the notations used in the model, their definitions, and units.  The model assumes a processing element, such as CPU, GPU, radio, or their combination with a dynamic power range, and batteries reside within an enclosure of a given size.  The processors, batteries, and the air within the enclosure each have an associated temperature ($T$), mass ($m$), and thermal capacity ($C$), which is the heat required to change the temperature of the mass, and is in units of joules per degree K.  We also assume the ambient temperature outside ($T_{amb}$) is unaffected by any heat transfer with the enclosure. 

The enclosure provides insulation from the environment based on its heat transfer coefficient $U$, which is an empirically derived constant that dictates the heat transfer rate ($\hat{Q}_{trans}$) in joules per unit time between the enclosure and its external environment. The overall heat transfer coefficient $U$ is a combination of the internal convection inside the enclosure, conduction through the enclosure walls, and external convection away from the enclosure. We can calculate $U$ as  thermal resistors connected in series, as below.

\vspace{-0.3cm}
\begin{equation}
\frac{1}{U} = \frac{1}{h_i} + \frac{d}{k} + \frac{1}{h_o}. 
\label{eq:series-transfer-coeff}
\end{equation}

In Equation~\ref{eq:series-transfer-coeff}, $h_i$ is the internal convection coefficient, $k$ is the thermal conductivity, $d$ is the thickness of the enclosure, and $h_o$ is the outer convection coefficient. The heat transfer rate ($\hat{Q}_{trans}$) below between the enclosure and its external environment is directly proportional to the heat transfer coefficient ($U$), temperature difference ($\Delta T(t)$) between inside and outside the enclosure, and the heat transfer area ($A$), computed as below.

\vspace{-0.4cm}
\begin{equation}
\hat{Q}_{trans} = U \times A \times \Delta T(t) = U \times A \times (T_{amb} - T_{enc}(t)).
\label{eq:q-trans-hat}
\end{equation}

For simplicity, our model assumes the enclosure is a cube with side length $L$ with a surface area $A=6L^2$.  As shown in Equation~\ref{eq:q-trans-hat}, the heat transfer rate is a function of surface area $A$.  At any time $t$, $\Delta T(t)=T_{amb} - T_{enc}(t)$ represents the difference between the temperature inside and outside the enclosure. Thus, a positive $\Delta T(t)$ represents heat flowing into the enclosure, and a negative $\Delta T(t)$ represents heat flowing out of it.  
Given Equation~\ref{eq:q-trans-hat}, we can compute the total heat transfer (in joules) over a time interval $\Delta t$ by simply integrating $\hat{Q}_{trans}$ over time, calculated as below.

\vspace{-0.3cm}
\begin{equation}
Q_{trans} =  \int_{t}^{t+\Delta t} \hat{Q}_{trans} \,dt.
\label{heat_transfer} 
\end{equation}

\begin{table}[t]
    \footnotesize
    \begin{center}
    \begin{tabular}{|| l | l | l ||} \hline
    {\bf Notation} & {\bf Definition} & {\bf Unit} \\ \hline \hline  
    { Q} & Heat transfer rate & J.s$^{-1}$ \\ \hline
    { h} & Heat transfer coefficient & W.m$^{-2}$K$^{-1}$ \\ \hline
    { U} & Combined heat transfer coefficient & W.m$^{-2}$K$^{-1}$ \\ \hline
    { k} & Thermal conductivity & W.m$^{-1}$K$^{-1}$ \\ \hline
    { d} & Enclosure thickness & m \\ \hline
    { A} & Enclosure surface area & m$^2$ \\ \hline
    { T} & Temperature & K \\ \hline
    { $\Delta$T} & Temperature difference & K \\ \hline
    { m} & Mass & Kg \\ \hline
    { C} & Thermal capacity & J.K$^{-1}$ \\ \hline
    { R$_{specific}$} & Specific gas constant & J.Kg$^{-1}$K$^{-1}$ \\ \hline
    { $p$} & Pressure & J.m$^{-3}$ \\ \hline
    { $\rho$} & Density & Kg.m$^{-3}$ \\ \hline
    { V} & Volume & m$^{3}$ \\ \hline
    { $\mathcal{V}$} & Processing element voltage & volts \\ 
    \hline
    { $\mathcal{I}$} & Processing element current & ampere \\ \hline
    { AF} & Airflow & m$^3$.s$^{-1}$ \\ \hline
    \end{tabular}
    \vspace{0.05cm}
    \caption{\emph{Model notations, definitions, and units.}}
    \label{tbl:notations}
    \end{center}
    \vspace{-1.1cm}
\end{table}

Equation~\ref{heat_transfer} enables us to compute the total heat energy transferred between the inside of the enclosure and the outside environment.  However, some of this heat energy is absorbed by the mass within the enclosure, including the processors, batteries, and air, and thus does not contribute to raising the enclosure's temperature ($T_{enc}$).  This heat energy is a function of the enclosure's heat capacity ($C_{enc}$), which is computed as the mass-weighted average of the respective heat capacities of the objects within the enclosure, as shown below, where $m_{enc} = m_{air} + m_{bat} + m_{proc}$ or the total mass within the enclosure.  Here, we assume the enclosure includes only processing elements, batteries, and ambient air.

\vspace{-0.35cm}
\begin{equation*}
C_{enc} = \frac{1}{m_{enc}}\times (m_{air}.C_{air} + m_{bat}.C_{bat} + m_{proc}.C_{proc} )  
\end{equation*}

\begin{figure*}[t]
    \centering
    \begin{tabular}{ccc}
    \includegraphics[width=0.3\textwidth]{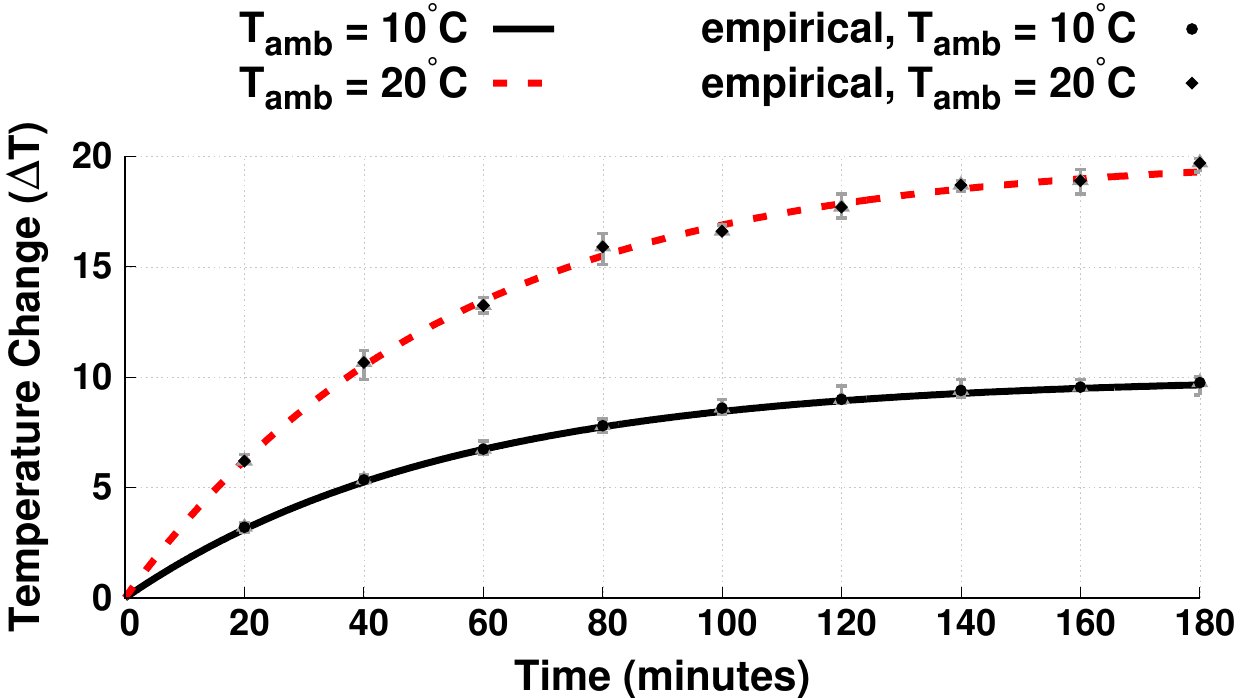} &
    \includegraphics[width=0.3\textwidth]{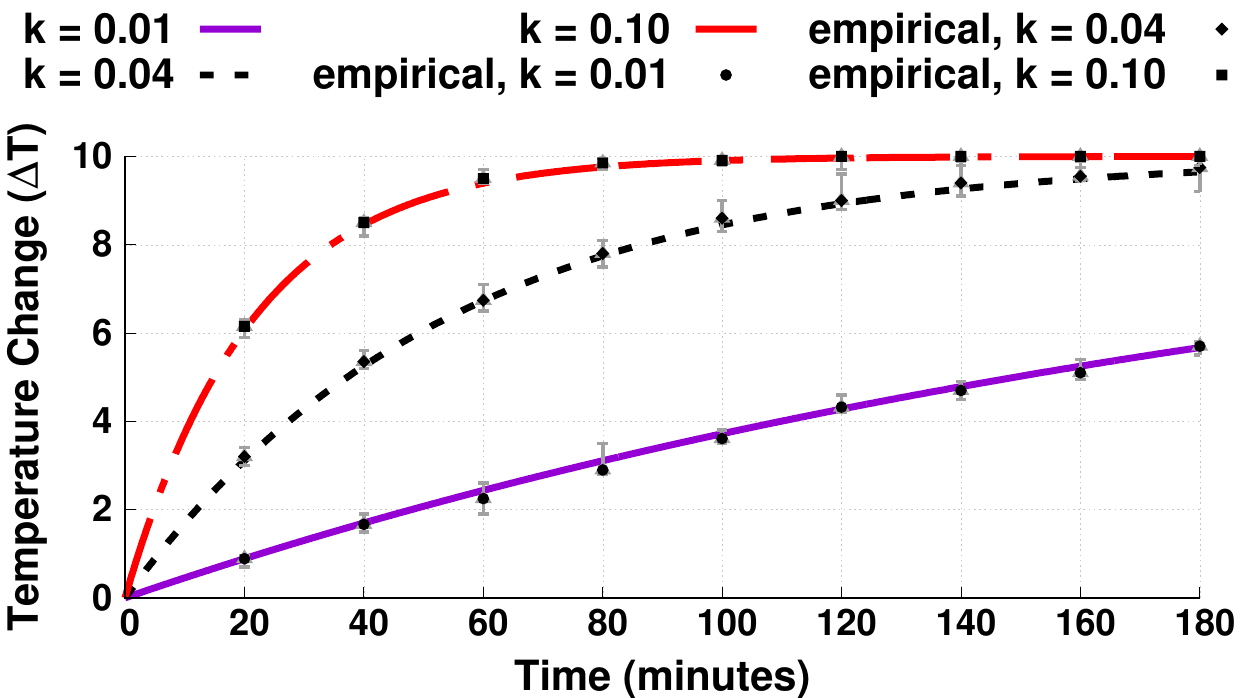} &
    \includegraphics[width=0.3\textwidth]{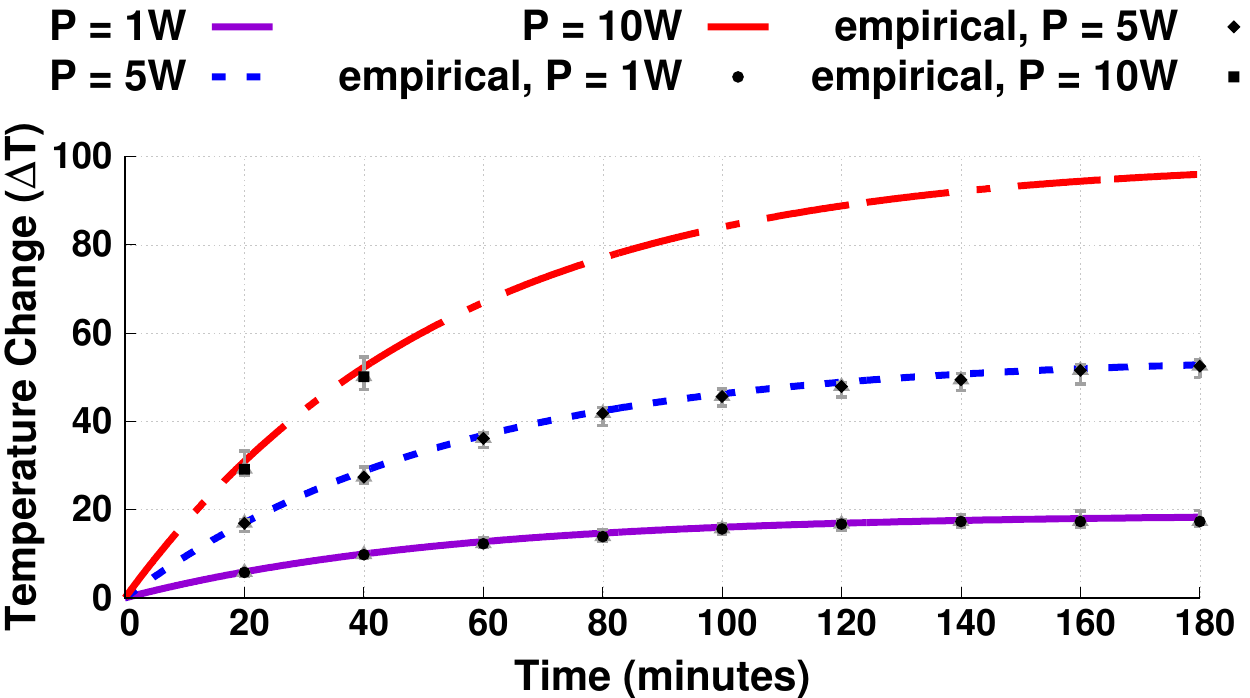} \vspace{-0.0cm}\\
    (a) Ambient Temperature ($k$=$0.04$, $P$=$0$)&
    (b) Thermal Conductivity ($T_{amb}$=$10$$^{\circ}C$, $P$=$0$)&
    (c) Power Usage ($k$=$0.04$, $T_{amb}$=$10$$^{\circ}C$) \\
    \end{tabular}
    \vspace{-0.35cm}
    \caption{\emph{The change in the enclosure's temperature $T_{enc}$ over time is a function of the (a) ambient temperature $T_{amb}$, (b) the enclosure's thermal conductivity $k$, and (c) the processor's power usage $P$.}}
    \vspace{-0.5cm}
    \label{fig:model_verification}
\end{figure*}

We can empirically measure the battery and computing platform's mass ($m$) and heat capacity ($C$) using a scale and calorimeter, respectively.  We cannot directly weigh the air mass, but can derive it using simple models. In particular, the mass of air in the enclosure is a product of its volume and the air's density $\rho_{air}$, which is directly proportional to the atmospheric pressure, and inversely proportional to the temperature ($T_{air}$) as well as its specific gas constant ($R_{specific}$), as given below.

\vspace{-0.2cm}
\begin{equation}
\rho_{air} = \frac{p}{R_{specific} \times T_{air}}.
\end{equation}

For dry air on earth $R_{specific} = 287.058 J \cdot kg^{-1} \cdot K^{-1}$. 
We assume the enclosure is closed when $p = 1$ atmosphere and $T_{air}=25^\circ$C.  
Note that, based on the ideal gas law, even when the temperature inside the enclosure changes, the ratio of its pressure $p$ to its temperature $T_{air}$, and thus its air density, remains constant.   As a result, in this case, the density of air $\rho_{air}=1.1839$kg/m$^3$, which results in an air mass $m_{air} = 1.1839 \times L^3$.  The heat capacity of air ($C_{air}$) at earth's surface under these conditions is also a constant and equal to 717 joules per kilogram per degree Kelvin (K). 
We have retrieved the coefficients for the thermal properties of the air and other components from Engineering ToolBox~\cite{air-thermal}.

Given all this, we can compute the enclosure's heat capacity $C_{enc}$ above. If the enclosure generates no internal heat, then its temperature will eventually reach an equilibrium temperature equal to the temperature $T_{amb}$ of the ambient environment.  
To reach equilibrium, the total amount of heat $Q_{trans}$ the enclosure will absorb or release is the product of its total mass $m_{enc}$, heat capacity $C_{enc}$, and change in temperature, computed as below.

\vspace{-0.3cm}
\begin{equation}
Q_{trans} = m_{enc} \times C_{enc} \times (T_{amb} - T_{enc}(0)).
\end{equation}

Here, $T_{enc}(0)$ is enclosure temperature at the start. While the equation above represents the heat transferred to reach the equilibrium temperature, the same basic equation also dictates the heat transferred over any arbitrary time interval $\Delta t$ based on the change in temperature at the time interval's start and end, as given below.

\vspace{-0.3cm}
\begin{equation}
Q_{trans} = m_{enc} \times C_{enc} \times (T_{enc}(t+\Delta t) - T_{enc}(t)).  
\label{capacity}  
\end{equation}

Notice that we have computed the total heat transfer $Q_{trans}$ over a time interval in both Equation~\ref{heat_transfer} and Equation~\ref{capacity}.   Setting these equations equal to each other yields our model, which predicts the temperature within the enclosure after some time interval $\Delta t$ given a starting temperature $T_{enc}(t)$, the enclosure's mass ($m_{enc}$) and heat capacity ($C_{enc}$), as well as its thermal conductivity ($k$), surface area ($A$), ambient temperature ($T_{amb}$), and depth ($d$). 

\vspace{-0.4cm}
\[ m_{enc} \times C_{enc} \times (T_{enc}(t+\Delta t) - T_{enc}(t))  = \int_{t}^{t+\Delta t} \hat{Q}_{trans} \,dt \]

\begin{equation}
T_{enc}(t+\Delta t)  = T_{enc}(t) +  \frac{1}{m_{enc}C_{enc}} \int_{t}^{t+\Delta t} \hat{Q}_{trans} \,dt
\label{capacity2}  
\end{equation}

To this point, our model assumes the processor generates no heat.  In practice, however, the power drawn by the processor is converted to heat, which our model assumes is uniformly distributed throughout the enclosure.  For now, we assume there are no mechanical components, such as fans, to dissipate this heat. 
We discuss modeling for heat dissipation using a fan below.  
Thus, we extend our model above to account for the processors' power draw by assuming it is entirely converted to heat.  We can account for this heat energy by simply adding it to the heat transferred with the environment, given as below.

\vspace{-0.4cm}
\begin{equation*}
T_{enc}(t+\Delta t)  = T_{enc}(t) +  \frac{1}{m_{enc}C_{enc}} \int_{t}^{t+\Delta t} (\hat{Q}_{trans} + (\mathcal{V} \cdot \mathcal{I})) \,dt.
\label{capacity3}  
\end{equation*}

Here, $\mathcal{V}$ and $\mathcal{I}$ are platform's voltage and current, and $\mathcal{V} \cdot \mathcal{I}$ is its overall power draw.  The model above simply observes that power translates directly to heat within the enclosure and thus augments any other heat transfer mechanism available to the system. 

Our model captures an enclosure's temperature change over time based on its physical characteristics, the ambient temperature, and the processor's power usage.  Unfortunately, there are no good physical models that capture the effect of temperature and power draw on usable battery capacity. Thus, we use empirical models from our battery's datasheet in Figure~\ref{fig:discharge_characteristics}, which we experimentally validated. We next extend our model to include using an arbitrary cooling component, such as an air conditioner or simple fan. 

\noindent{\bf Heat Dissipation Using Active Cooling.} Processors may dissipate heat faster than an enclosure can transfer it to the external environment based on its conductivity, which causes the temperature to rise to a point where neither the battery nor processor can function.  
In this case, fans or pumps may be necessary to increase the rate of heat dissipation within the enclosure using convection.  
While our model below focuses on fans, which transfer heat by moving air, the same basic approach applies to pumps, which transfer heat by moving a liquid. 
The selection of a fan depends on the specifications of the enclosure and the maximum rate of heat dissipation required. A fan moves the air that absorbs the heat from inside the box and then dissipates it to the external environment. The amount of energy dissipated ($Q_{diss}$) depends on the mass of the moving air ($m_{air}$), the specific heat of the moving air ($C_{air}$), and the temperature change of the moving air ($\Delta T_{air}$).

\vspace{-0.3cm}
\begin{equation*}
Q_{diss}  = m_{air} \times  C_{air} \times \Delta T_{air}
\label{q_diss1}  
\end{equation*}

The mass of the moving air can be calculated from the volume of air ($V_{air}$) being moved and the density of the moving air ($\rho_{air}$).

\vspace{-0.3cm}
\begin{equation*}
Q_{diss}  = (V_{air} \times \rho_{air}) \times  C_{air} \times \Delta T_{air}
\label{q_diss2}  
\end{equation*}

We divide both sides of the equation to get the power needed to dissipate at each time ($t$).

\vspace{-0.2cm}
\begin{equation*}
\frac{Q_{diss}}{t}  = (\frac{V_{air}}{t} \times \rho_{air}) \times  C_{air} \times \Delta T_{air}
\label{q_diss3}  
\end{equation*}

The air volume over time is the air flow rate, which we term as $AF_{air}$. We term the power dissipation as $P_{diss}$. We arrange the equation to get the airflow required for a given power dissipation.

\vspace{-0.3cm}
\begin{equation*}
AF_{air}   = \frac{P_{diss}}{\rho_{air} \times  C_{air} \times \Delta T_{air}}
\label{q_diss4}  
\end{equation*}

The value of $C_{air}$ is $1 kJ \cdot kg^{-1} \cdot C^{-1}$ and the density of air at $20^\circ$C is  $\rho_{air}=1.20$kg/m$^3$. We can use this equation to either find the heat dissipation rate for a given flow rate or the heat flow required to achieve the desired heat dissipation rate. 

Since the equipment inside the enclosure will exhibit resistance to the air flow, air needs to be delivered at a certain pressure that can overcome the resistance. However, the amount of pressure required is highly dependent on the design and physical characteristics of the product to be cooled, and must be determined either experimentally using anemometers and manometers to measure the air speed and pressure, respectively, or using different computer-aided design (CAD) software to design and calculate airflow characteristics. Either method will yield system pressure requirements that increase with the air flow. 
The pressure exerted by the fan reduces as its air flow increases, and delivers the highest air flow when the back pressure is lowest.
The intersection of the two curves is the operating point of the fan for the given system. We assume our model uses a fan that has the required airflow at its operating point. The power consumption of a variable speed fan has a cubic relationship with the change in the airflow. That is, if the airflow of the fan doubles, its power consumption increase by 8$\times$. This relationship is shown in the equation below.

\vspace{-0.4cm}
\begin{equation*}
P_j = P_{i} \times (\sfrac{AF_j}{AF_i})^3 
\label{fan_power_law}  
\end{equation*}
\vspace{-0.4cm}

Here, $P_i$ and $AF_i$ are the initial power consumption and airflow while $P_j$ and $AF_j$ are the final power consumption and airflow, respectively. Note that using the fan to dissipate heat reduces the energy available for doing productive computation. 

\begin{figure}[t]
    \includegraphics[width=0.26\textwidth]{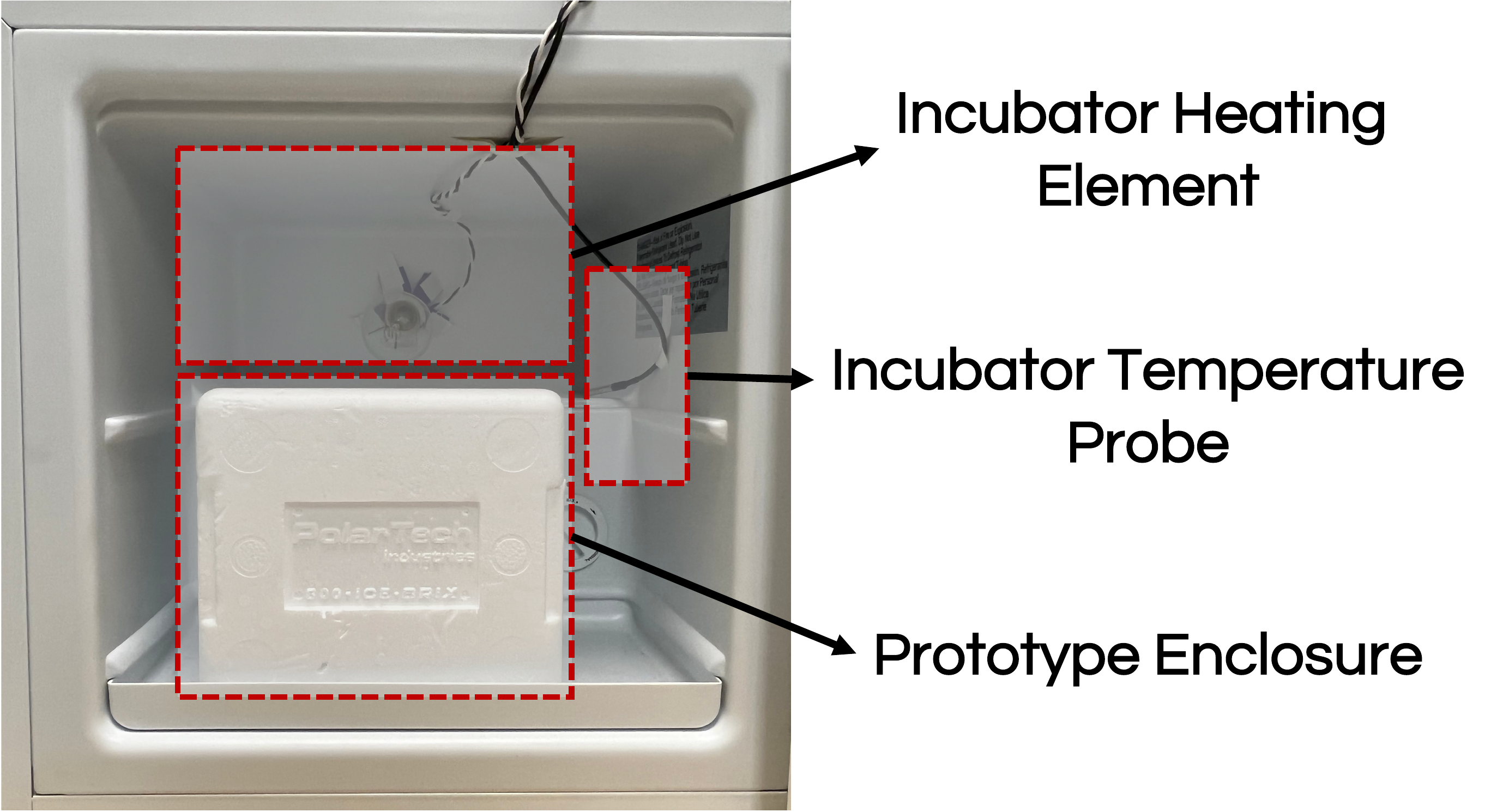}
    \vspace{-0.35cm}
    \caption{\emph{Small-scale prototype inside the incubator.}}
    \label{fig:incubator}
    \vspace{-0.7cm}
\end{figure}

\noindent{\bf Implementation and Model Validation.}
To validate our model and experiment with thermodynamic design, we built a programmable incubator by connecting a mini-freezer and incandescent light bulb (as a heat source) to programmable relays controlled by a Raspberry Pi (Figure~\ref{fig:incubator}).  Our incubator programmatically controls temperature between -30$^\circ$C and 40$^\circ$C with an error of $\pm$$2$$^\circ$C. We use the Nvidia Jetson Nano as our computing platform for validation.  The Nano has a baseload power of $\sim$1W, and a maximum power of 10W.  We use Panasonic NCR18650B lithium-ion batteries rated for 3.2 Amp-hours (Ah) at 3.6V.  We use a boost converter to build a 4Ah, 5V battery bank for the Nano with 20Wh maximum energy capacity. Figure~\ref{fig:discharge_characteristics} from \Section\ref{sec:background} shows our battery's response to temperature and discharge current.  Our enclosure uses Expanded Polystyrene (EPS) foam, which has a thermal conductivity $k$ of 0.04$Wm^{-1}K^{-1}$~\cite{eps-conductivity}. We use 1.25in thickness as our baseline for experiments.  We vary $k$ by changing its thickness. For example, halving the thickness increases $k$ by 2$\times$ to 0.08$Wm^{-1}K^{-1}$.

Three parameters affect the change in enclosure's temperature $T_{enc}$ over time: the difference with the ambient temperature $T_{amb}$, the enclosure's thermal conductivity $k$, and the processor's power usage $P$. Figure~\ref{fig:model_verification} shows the complex non-linear effect of each on the change in $T_{enc}$, initialized to 0C.  Our baseline is $T_{amb}$=$10^\circ$C, $k$=$0.04$, and $P$=$0$W. Figure~\ref{fig:model_verification}(a) then varies $T_{amb}$ without changing $k$ or $P$, and shows that a higher ambient temperature causes the enclosure's temperature to rise more quickly.  Similarly, Figure~\ref{fig:model_verification}(b) varies $k$, and shows that a higher thermal conductivity, i.e., more insulation, also causes the temperature to rise quickly. Finally, Figure~\ref{fig:model_verification}(c) varies the processor's power usage, and shows how the resulting heat increases the temperature up to $100^\circ$C at full utilization (10W), which is well beyond the $10^\circ$C ambient temperature.  Figure~\ref{fig:model_verification} also validates our model: the curves represent our model, while datapoints represent the mean temperature change across five experiments, where error bars represent the max and min.

\section{Thermodynamic Model Use Cases}
\label{sec:use-cases}
\begin{figure}[t]
    \includegraphics[width=0.85\linewidth]{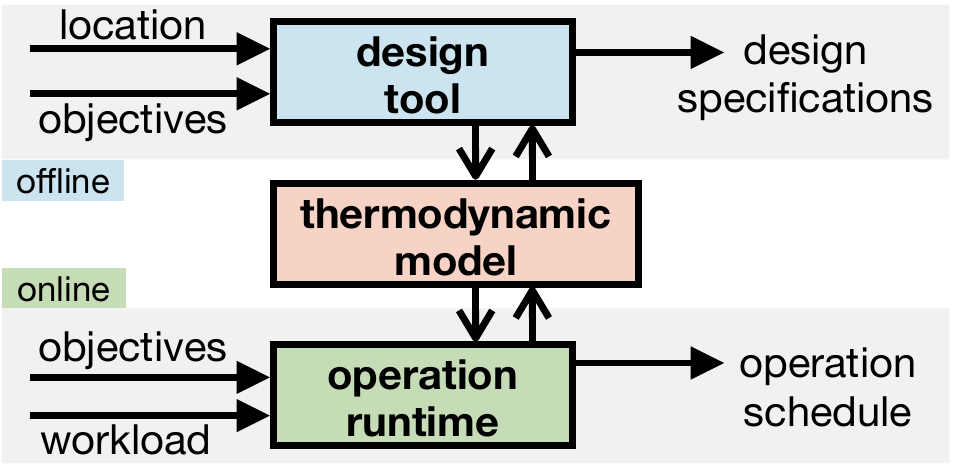}
    \vspace{-0.3cm}
    \caption{\emph{An overview of thermodynamic model use cases.}}
    \label{fig:use-cases}
    \vspace{-0.65cm}
\end{figure}

In this section, we first present the design of an environmentally-powered computer system, specifically its enclosure, and operation of the system as the two uses cases for the thermodynamic model, as shown in Figure~\ref{fig:use-cases}. 
We then present a broader analysis that outlines the use cases across a wide range of settings.

\vspace{-0.2cm}
\subsection{Use Case 1: Designing the System}
\label{sec:enclosure_design}
\vspace{-0.1cm}
In the design use case, the end goal is to determine the configuration range for the system enclosure that allows the system to meet its performance objectives across seasons. 
To do so, a pre-requisite is the historical temperature profile of the system's location and user-specified system objectives, such as 100\% availability at 50\% of the power. User also specifies the order of priority for secondary metrics.
In addition, user must also specify the capacity of different system components such as the processor, batteries, and solar panels. 
Given these inputs, we exhaustively search the system enclosure parameter space, which includes the enclosure insulation and cooling element capacity, to find a range of parameters that satisfy the user-specified performance objectives. 

We take an iterative approach to finding the right specifications for the enclosure and the cooling element. 
We start with an initial set of values for the enclosure's insulation, or thermal conductivity $k$, that may correspond to low (e.g., styrofoam, $k=0.04W/m.K$), medium (e.g., Polyvinyl Chloride (PVC) plastic $k=0.2W/m.K$), and high (e.g., glass, $k=0.8W/m.K$) value. 
These values are configurable and correspond to actual insulation materials that can be used for the enclosure. 
Similarly, we pick an initial value for the cooling capacity, specified in watts (W). We then use an iterative process to find the optimal enclosure and cooling element specifications using our thermodynamic model based on the temperature and solar power profiles at all insulation values.
Since the workload  is not known, we vary the system's operational power between the user-specified minimum (e.g., 30\%) and 100\% power. 
For each of the configuration combinations (i.e., thermal conductivity, cooling capacity, and operational power), we compute the value of all of the metrics that quantify system performance objectives. 

We then use a brute-force approach to find the best  configurations.  Since finding the right enclosure specification is a one time process done before system deployment, the time required for our brute-force approach is not a problem. Finally, we output a single configuration that satisfies the primary performance objective while maximizing the other metrics in the order of their priority.  It is possible that no configuration meets the desired level of performance for the provided specifications. 

\begin{figure}[t]
    \includegraphics[width=0.85\linewidth]{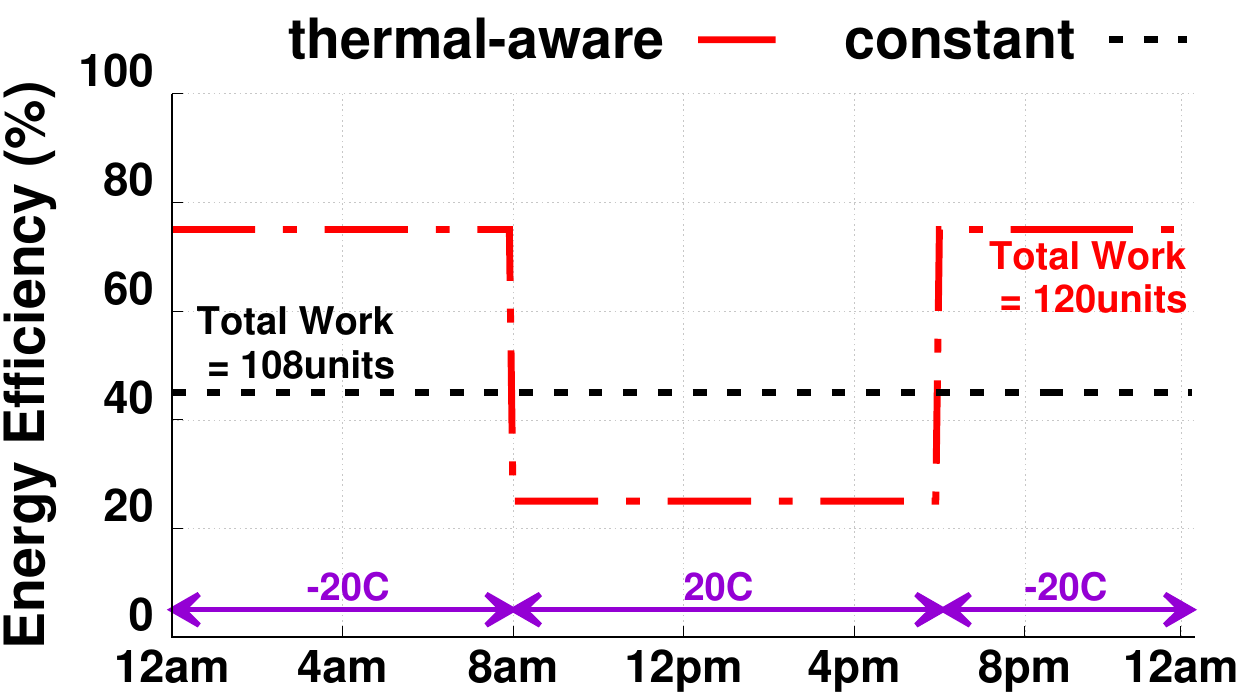}
    \vspace{-0.35cm}
    \caption{\emph{Work done using i) naive and ii) thermal-aware scheduling that exploits the electrical/thermal feedback loop.}}
    \label{fig:scheduling-energy-effect}
    \vspace{-0.65cm}
\end{figure}

\vspace{-0.2cm}
\subsection{Use Case 2: Operating System Components}
\label{sec:enclosure_operation}
\vspace{-0.1cm}
In the operation use case, our goal is to determine the operating point of the computing and cooling elements that allows the system to meet its performance objectives over a finite scheduling horizon. 
To do so, we require all the inputs of the design use case and the output parameters of the design process with one key distinction. 
Instead of the observed temperature, solar generation, and workload arrival schedule, we need forecasted values for these inputs. 
Temperature forecasts are generally highly accurate and readily available.
Solar power forecasts are also available through many open-source and publicly-available tools, such as Solar-TK~\cite{solar-tk}.
The workload patterns for applications that are deployed using small-scale embedded systems or edge datacenters tend to also be deterministic. 
Given these inputs, we simulate our thermodynamic model using the forecasted values of temperature, solar power, and workload. In doing so, we schedule the workload for each hour of the scheduling horizon, while satisfying the specified objectives.

In determining the operational schedule, we leverage the insight that thermal and electrical energy both exhibit a feedback loop between themselves, and that the thermal energy generated due to energy consumption at time $t$ affects the availability of energy in subsequent time periods. We use a simple example to demonstrate this effect, which we term the \emph{scheduling-energy effect}. 
It refers to the relationship between the intensity at which the processors operate, and the energy available from the battery.  
Specifically, due to the temperature-energy effect described in Section~\ref{sec:background}, the warmer the battery, the more energy a system can extract from it.  
Thus, when scheduling workload, systems can extract more energy, and perform more total computation, if they operate at a higher utilization as the temperature decreases, and lower utilizations when the temperature increases.  
The former maintains a higher temperature, which increases battery efficiency, while the latter generates less heat, which reduces the need for cooling, which consumes additional energy to dissipate the waste heat.  

\begin{figure}[t]
    \includegraphics[width=0.8\linewidth]{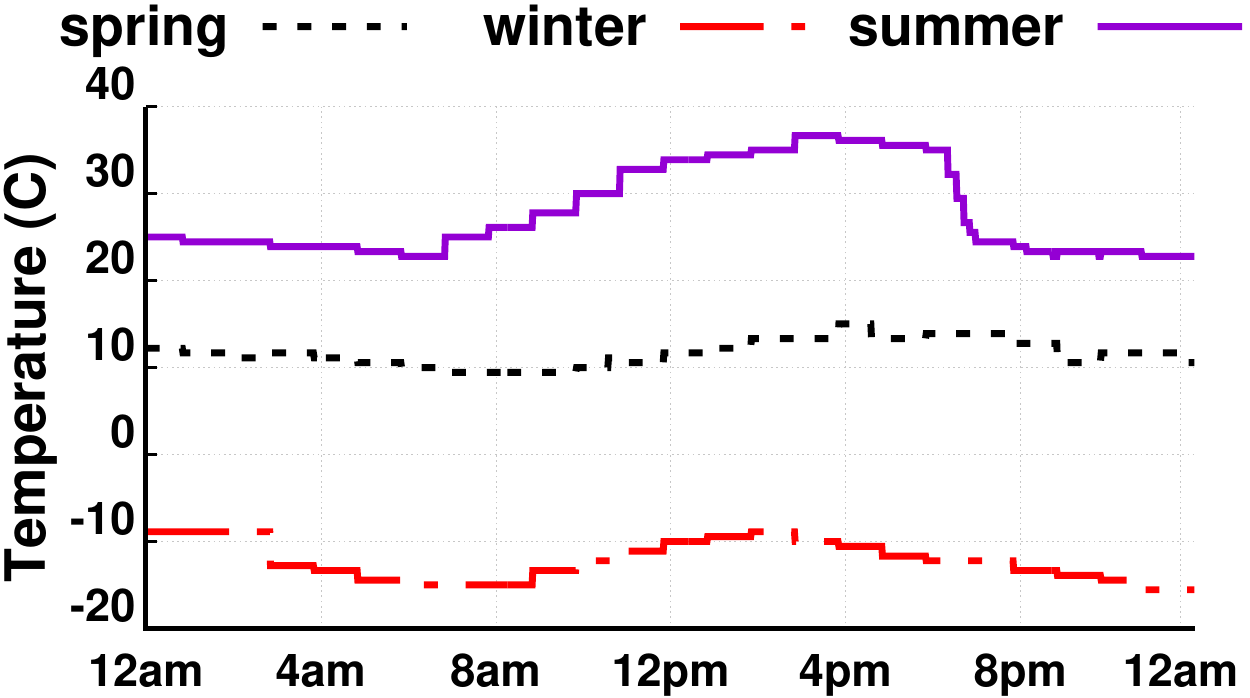}
    \vspace{-0.3cm}
    \caption{\emph{Temperature profiles of a single location in northeast United States across different seasons.
    }}
    \label{fig:temperature-profiles}
    \vspace{-0.6cm}
\end{figure}

Figure~\ref{fig:scheduling-energy-effect} quantifies the scheduling-energy effect in a scenario where the temperature drops over night but rises during the day.   
The graph compares operating continuously at $\sim$50\% utilization with a scheduling policy that operates at $\sim$25\% utilization during the day and $\sim$80\% utilization over night.   
In this case, the latter schedule is able to perform 11\% more computation than the former because of the effects above.  
For this experiment, our computations is simply an integer benchmark.  
The experiment demonstrates that scheduling  \emph{when} and \emph{how much} processors dissipate heat can affect a system's energy-efficiency and the total computation. 

We use a simple iterative approach to find an optimal workload schedule that meets the user's performance objective.
We work backwards from the end of the scheduling horizon and schedule workload for each hour such that the energy is extracted from the battery at the highest energy-efficiency. In the first round, it simply uses the default workload pattern and then changes the workload in each slot to ensure the performance objective is met.

\vspace{-0.2cm}
\subsection{Thermodynamic Model at Work}
We next show how our thermodynamic model can improve the design and operation of the use cases above. 
We present multiple combinations of performance objectives and minimum operational power constraints. 
We also decouple the results across three seasons to demonstrate how the performance tradeoffs are impacted by seasonal variations even for a given location.

\vspace{-0.1cm}
\subsubsection{\bf Evaluation Setup}
Below, we outline the key metrics used to specify systems' performance objectives.

\noindent{\bf Performance Metrics.} 
We define system performance objectives using three  metrics: \emph{energy-efficiency}, \emph{availability}, and \emph{work rate}.

{ \emph{Energy-efficiency}} is defined as the percentage of available energy in the battery that is extracted and used for computation. 
\vspace{-0.05cm}
\begin{equation*}
    \text{Energy Efficiency} = 100 \times \frac{\text{Energy used for computation}}{\text{Energy stored in the battery}}
  \end{equation*}
\vspace{-0.05cm}
A value of 100\% means that all the energy extracted from the battery is used for the computation, while a value of 50\% means that only 50\% of the energy could be extracted and used for the computation. In this case, the other 50\% either could not be extracted from the battery or was used by the cooling system.  A higher value is better. 

\emph{Availability} is the percentage of time the system was up and has enough power to operate at or above a threshold utilization. 

\vspace{-0.3cm}
\begin{equation*}
    \text{Availability} = 100 \times \frac{\text{System up time}}{\text{Total experiment time}}
  \end{equation*}
\vspace{-0.3cm}

\begin{figure*}[t]
    \centering
    \begin{tabular}{ccc}
    \includegraphics[width=0.3\textwidth]{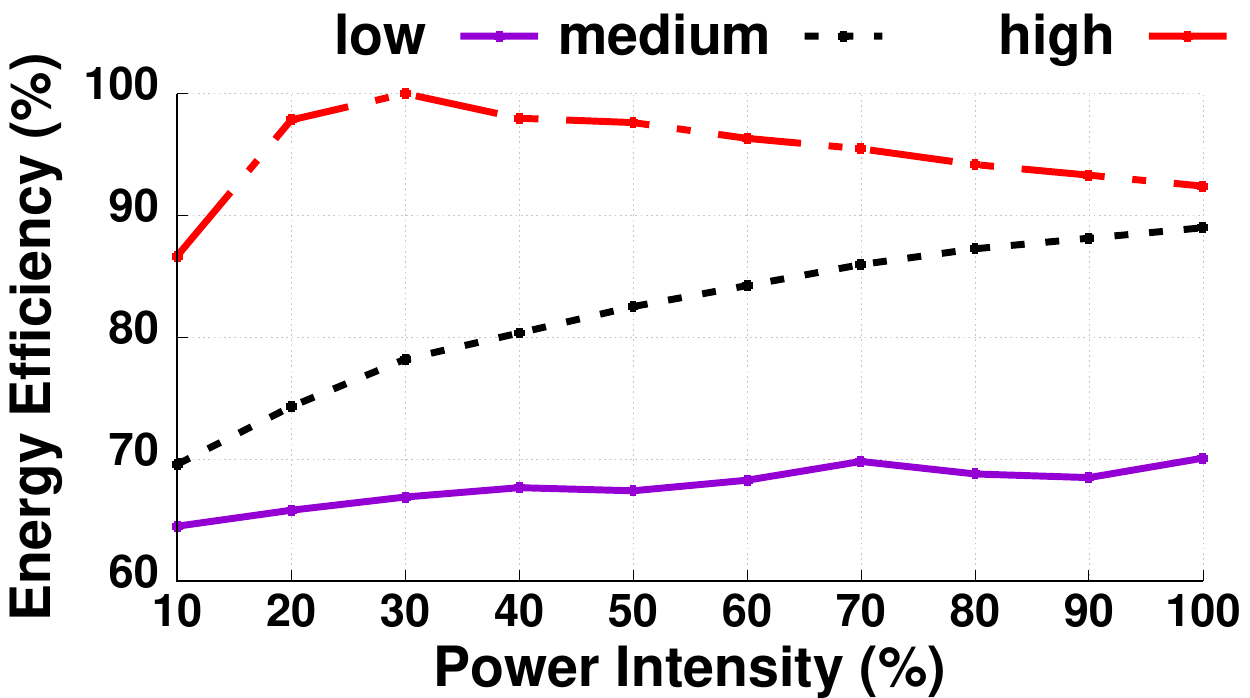} &
    \includegraphics[width=0.3\textwidth]{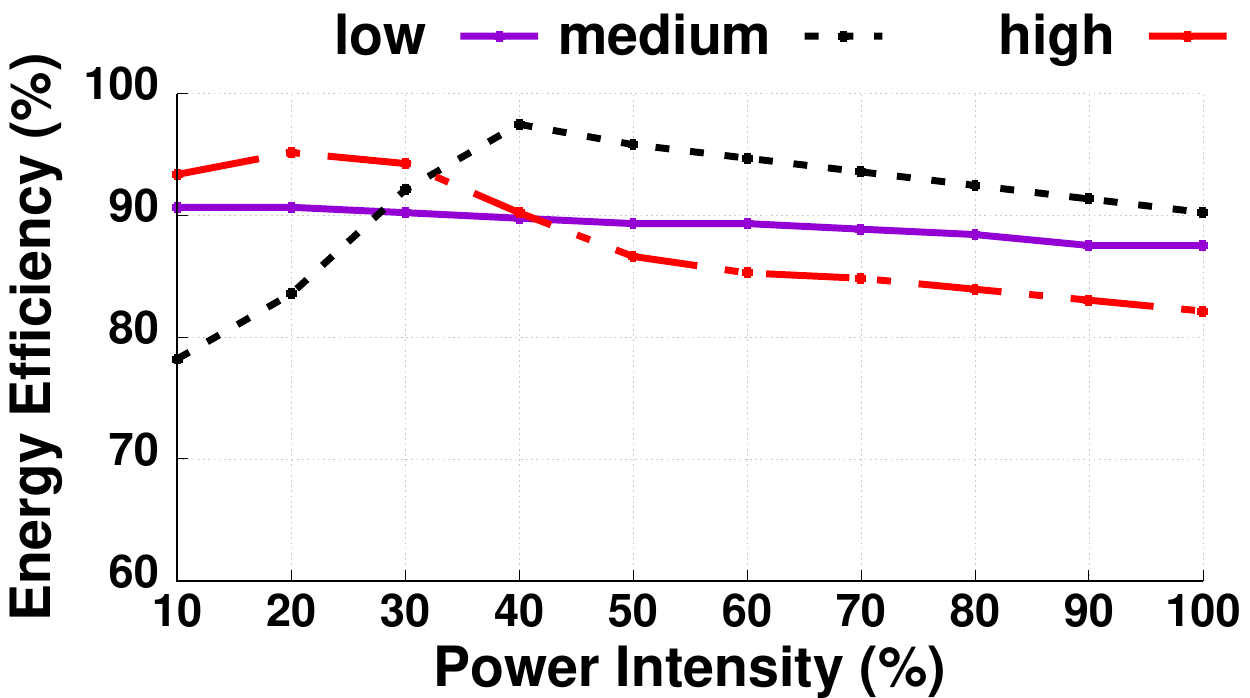} &
    \includegraphics[width=0.3\textwidth]{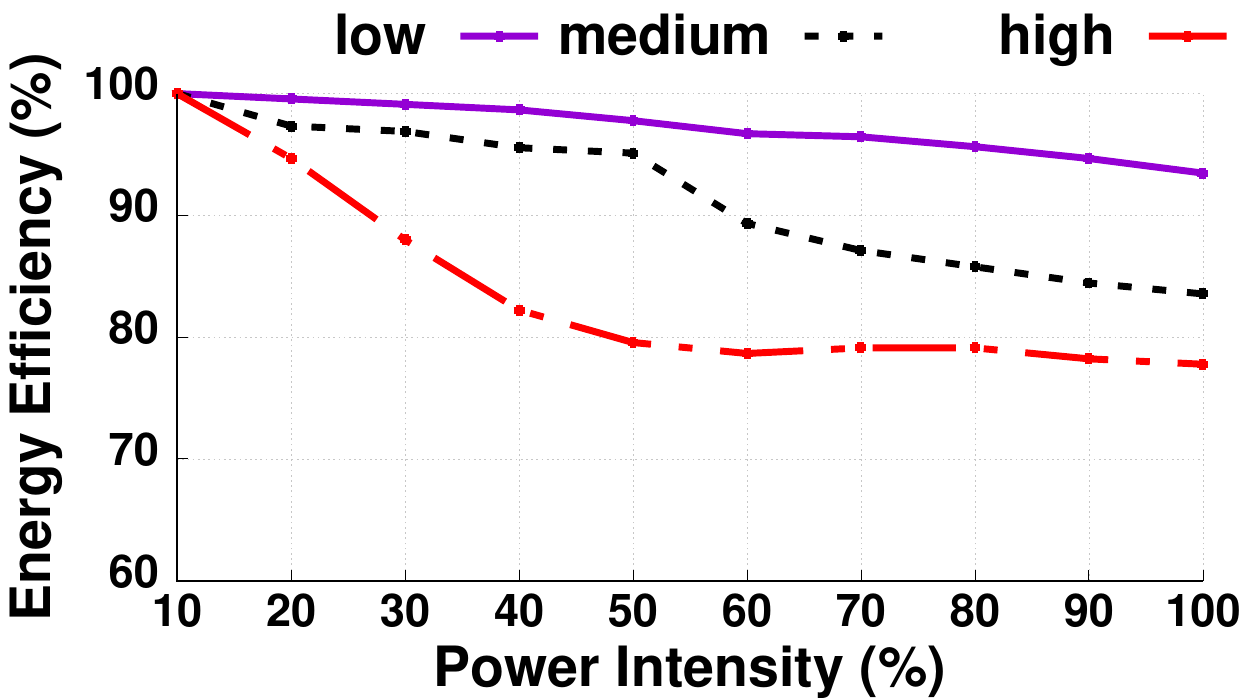}\\
    (a) Winter&
    (b) Spring&
    (c) Summer\\
    \end{tabular}
    \vspace{-0.4cm}
    \caption{\emph{Energy-efficiency across three seasons: (a) winter, (b) spring, and (c) summer. For each season, we evaluate our three designs that are winter-optimal (red line, long dash), spring-optimal (black line, small dash), and summer-optimal (purple line, solid) over various operating points represented by power intensity of x-axis.}}
    \label{fig:energy_efficiency}
    \vspace{-0.2cm}
\end{figure*}

\emph{Availability} ranges between 0\% and 100\%. A higher value is better for a given operating point. 
Finally, the \emph{work-rate} is defined as the amount of work done (computation) per unit time. Its value is in the $(0, \inf)$ range. Higher values of work rate are better.  

\begin{figure*}[t]
    \centering
    \begin{tabular}{ccc}
    \includegraphics[width=0.3\textwidth]{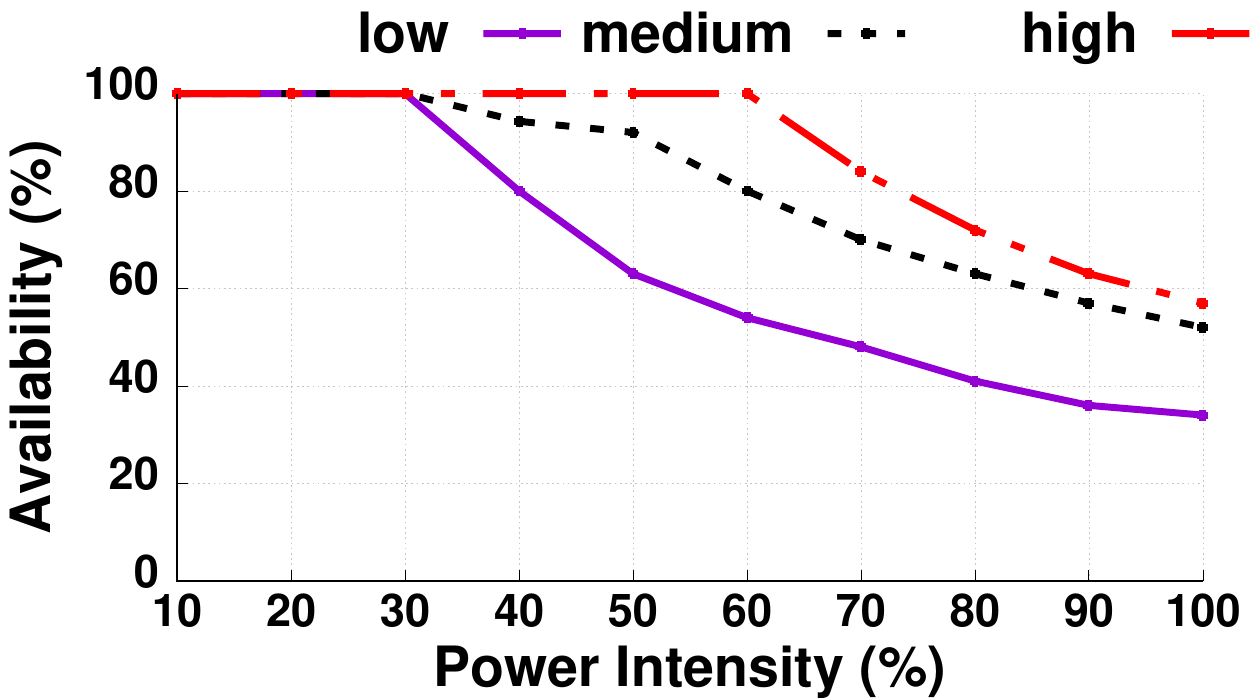} &
    \includegraphics[width=0.3\textwidth]{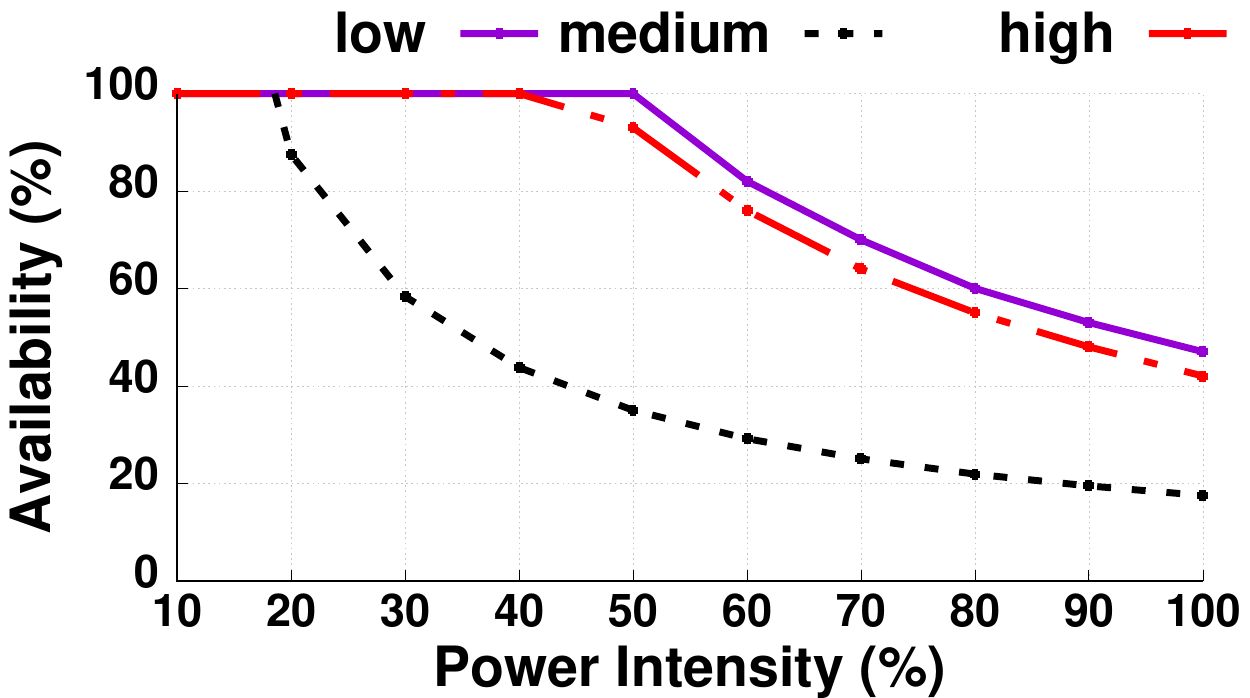} &
    \includegraphics[width=0.3\textwidth]{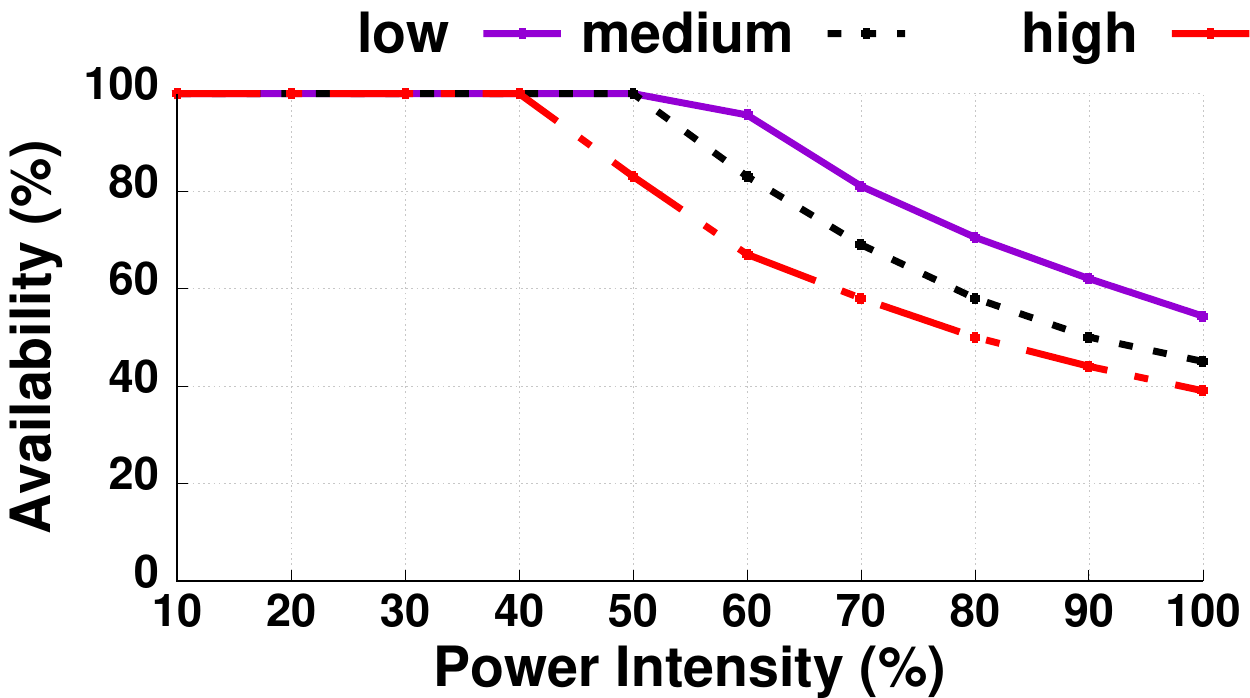}\\
    (a) Winter&
    (b) Spring&
    (c) Summer\\
    \end{tabular}
     \vspace{-0.4cm}
    \caption{\emph{Availability across three seasons: (a) winter, (b) spring, and (c) summer. For each season, we evaluate our three designs that are winter-optimal (red line, long dash), spring-optimal (black line, small dash), and summer-optimal (purple line, solid) over various operating points represented by power intensity of x-axis.}}
    \label{fig:availability}
    \vspace{-0.5cm}
\end{figure*}

\noindent{\bf System Configurations.}
For our use case demonstration, we consider a small-scale carbon-free edge datacenter similar to those considered in recent work~\cite{socc21,hotnets21}.   We consider the enclosure, in this case a small room, as a cube with 8ft sides. 
The size of the battery is 20kWh with a minimum state-of-charge of 40\% or 8kWh. 
The edge datacenter houses 8 servers of 250W each with a total demand of 2kW at 100\% utilization. The available battery capacity of 12kWh is enough to run all the servers at 50\% utilization for 24 hours under ideal conditions. 
This setting allows us to vary the system utilization (and the current draw) around 50\% and evaluate the effect of increasing or decreasing battery's discharge current on system's objectives.
The heat capacity, volume, and mass of the battery is based on our lithium-ion battery datasheet~\cite{battery-datasheet}. 

We demonstrate the effect of the design and operating point on the energy-efficiency, availability, and performance using a site in the northeast U.S. This site exhibits significantly different weather across winter, summer, and spring. The temperature profiles for three representative days of these seasons are shown in Figure~\ref{fig:temperature-profiles}. The temperature varies from 77$^\circ$F (25$^\circ$C) to 97$^\circ$F (36$^\circ$C) in summer, from 46$^\circ$F (8$^\circ$C) to 52$^\circ$F (11$^\circ$C) in spring, and 5$^\circ$F (-15$^\circ$C) to 16$^\circ$F (-5$^\circ$C) in winter. We configure the enclosure's heat transfer coefficient at three insulation settings: low (2), medium (0.9), and high (0.35). These values are achieved by varying the thickness of insulated wall with thermal conductivity of 0.15 W/m$\cdot$K. 

\vspace{-0.1cm}
\subsubsection{\bf Energy-efficiency}
Figure~\ref{fig:energy_efficiency} shows the energy-efficiency for different design parameters and operating points across all seasons. Here, we assume the system optimizes for energy-efficiency and discuss the design choices and operating points across seasons.

\noindent
\emph{\textbf{Effect of Design.}} 
The choice of design to optimize for energy-efficiency depends on which season the system optimizes for and how much loss of energy-efficiency it is willing to accept in other seasons. 
If energy-efficiency in winter is desired for the system, we should opt for a design that offers the highest protection against ambient weather and best performance in heat scavenging, termed as winter-optimal (high insulation). 
This design gives you 100\% energy efficiency in winter and offers the highest energy-efficiency for any operating point (Figure~\ref{fig:energy_efficiency}(a)). 
However, its energy-efficiency in other seasons is significantly lower, especially in summer (Figure~\ref{fig:energy_efficiency}(c)), since, in summer, this design has to use significant fan energy to dissipate the waste heat. 
Similarly, a low insulation design is the best choice for summer energy-efficiency. 
As expected, its performance in winter is the worst as high conductivity allows processor heat to escape, preventing it from retaining heat when idle. The spring-optimal design (medium insulation) offers the best performance in spring and fall (Figure~\ref{fig:energy_efficiency}(b)). Since its performance is better than low insulation in winter and high insulation in summer, it is the best choice to optimize energy-efficiency across seasons. 

\noindent
\emph{\textbf{Effect of Operation.}}
The design of a system for a season does not automatically guarantee the best energy-efficiency. 
The operating point, i.e., utilization, provides another knob that optimizes the energy-efficiency. For example, during winter, the energy-efficiency for the low insulation design is highest at the maximum operating point. 
This is because, at higher utilization, more heat is generated which keeps the battery warm. 
The gain in battery energy-efficiency is enough to offset the negative effect of higher current draw. 
However, the same design offers the best energy-efficiency at the lowest operating point in spring and summer. 
This is because, at low operating points, the low insulation is able to dissipate the heat through normal heat transfer. 
The low operating point not only avoids the use of a fan but also the negative impact of higher discharge currents. 
This trend is not same for all the design choices. 
For a high and medium insulation, the best performance is achieved at mid-operating points in their respective seasons. 
This demonstrates that, given a design, the choice of operating point will vary within a season and across seasons. 
Figure~\ref{fig:energy_efficiency} can also be used as a guide to designing systems. If the system must operate at a certain operating point, you can choose a design that gives the best performance. For example, if the system must always operate at 10\% power intensity, the high insulation gives the best performance both in winter and summer, and comparable performance for the rest of the year. 

\vspace{-0.1cm}
\subsubsection{\bf Availability}
Figure~\ref{fig:availability} evaluates availability for different design parameters and operating points across seasons. Here, we assume the system optimizes for availability and discusses the design choices and operating points to achieve that across seasons. 

\noindent
\emph{\textbf{Effect of Design.}} The availability across all operating points differs for each insulation level. The maximum availability offered by each insulation across all operating points differs significantly across seasons. A high insulation offers a minimum of 60\% availability across all operating points as compared to 35\% for the low insulation. 
This is because the energy-efficiency of the two designs varies significantly at the highest operating point. 
However, the same high insulation offers only 38\% availability at all operating points in summer. 
This is due to the energy loss to a fan, as it needs higher airflow to dissipate heat as the operating point increases. 
The choice of design for 100\% availability is straightforward if the operator does not care about the operating point or energy-efficiency. Figure~\ref{fig:availability} illustrates that all the design options offer 100\% availability across all seasons. They only differ by the highest operating point at which they offer 100\% availability. Thus, if the operator wants the system to be 100\% available, they can choose any design and then operate it at the highest operating point at which it offers 100\% availability.

\begin{figure}[t]
    \vspace{-0.05cm}
    \includegraphics[width=0.93\linewidth]{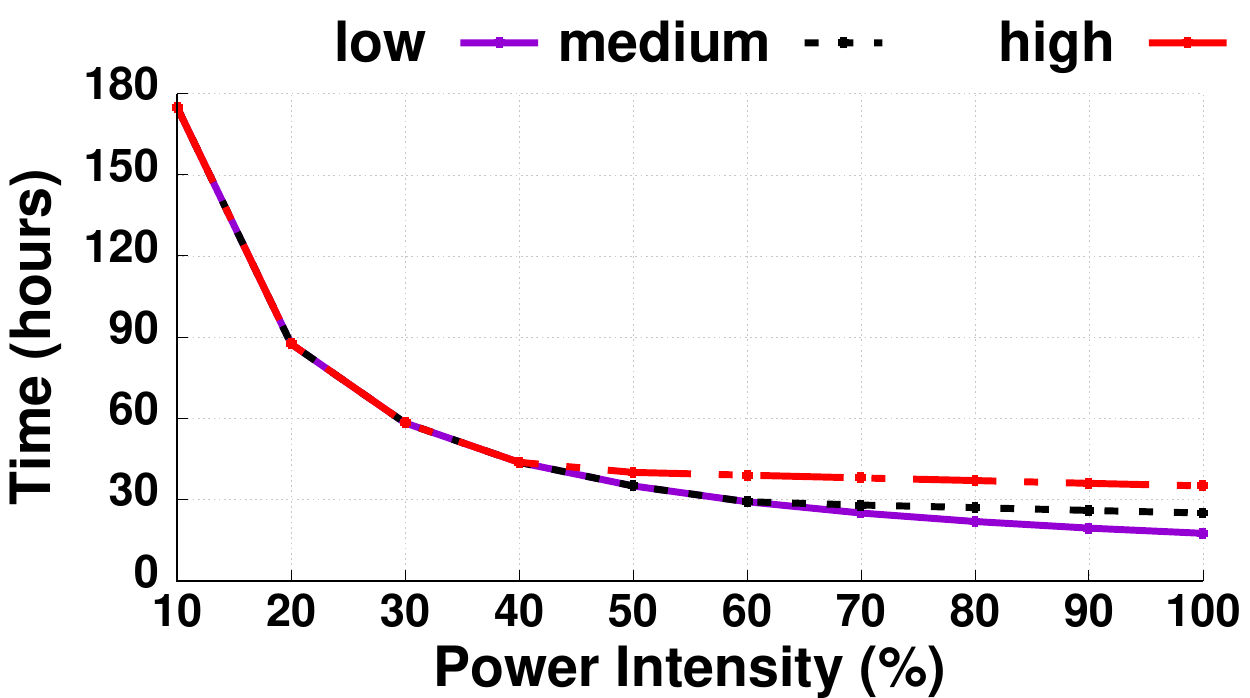}
    \vspace{-0.3cm}
    \caption{\emph{Performance during summer for high (winter-optimal), medium (spring-optimal), and low (summer-optimal) insulation designs over various operating points.}}
    \label{fig:performance}
    \vspace{-0.5cm}
\end{figure}

\noindent
\emph{\textbf{Effect of Operation.}}
Each operating point offers different availability across seasons.  For example, at 30\% or lower power intensity, the system achieves 100\% availability irrespective of design. 
This is less than the 50\% power intensity that an ideal system can support with 100\% availability. 
It shows the poor thermal management of designs for non-optimal seasons. 
Note that each design gives 100\% availability at a higher operating point in its optimal season. 
For example, a high insulation manages thermal energy the best in winter, as it exceeds the 50\% operating point. 
This is due to heat retention that takes the battery temperature above 25$^\circ$C and extracts more energy than the nominal value. This effect is consistent across seasons: the design that best manages thermal energy in a given season, achieves the highest operating point for 100\% availability. 

\vspace{-0.2cm}
\subsubsection{\bf Performance}
Figure~\ref{fig:performance} shows the performance of different designs during the summer. There are two key points in this evaluation. First, at different operating points, the speed of work differs. This is intuitive as power intensity on the x-axis is directly proportional to the CPU utilization. 
At 100\% utilization, the rate of computation is 10 times faster than at 10\% utilization. The second takeaway is the difference in performance across designs. 
The low insulation performs the best as it does not need a fan or active cooling. 
However, the other two designs must dissipate energy using a fan, as well as use workload scheduling to reduce the energy consumption of the fan. For example, the high insulation design uses a very simple scheduling policy to stop computation when the temperature exceeds 60$^\circ$C and resumes it only when the temperature drops to 60$^\circ$C using a combination of passive cooling through conduction and active cooling using the fan. This way, the design is able to achieve higher energy-efficiency at the cost of performance.

\section{Case Studies}
\vspace{-0.05cm}
\label{sec:case-studies}
We next present two applications as case studies that make use of our thermodynamic model while prioritizing different objectives. 
The first application is precision agriculture where an IoT base station is gathering sensor data from environmental sensors that are part of a distributed wireless sensor network. 
The state of the art for this application is Farmbeats~\cite{farmbeats}, which is an IoT platform for data-driven agriculture.
The second application is federated learning in smart cities where multiple edge computing platforms are training a machine learning (ML) model.  
These case studies show that our \emph{thermal-aware} design and operation achieves better performance  than application-specific state-of-the-art.

\begin{figure}[t]
    \includegraphics[width=0.91\linewidth]{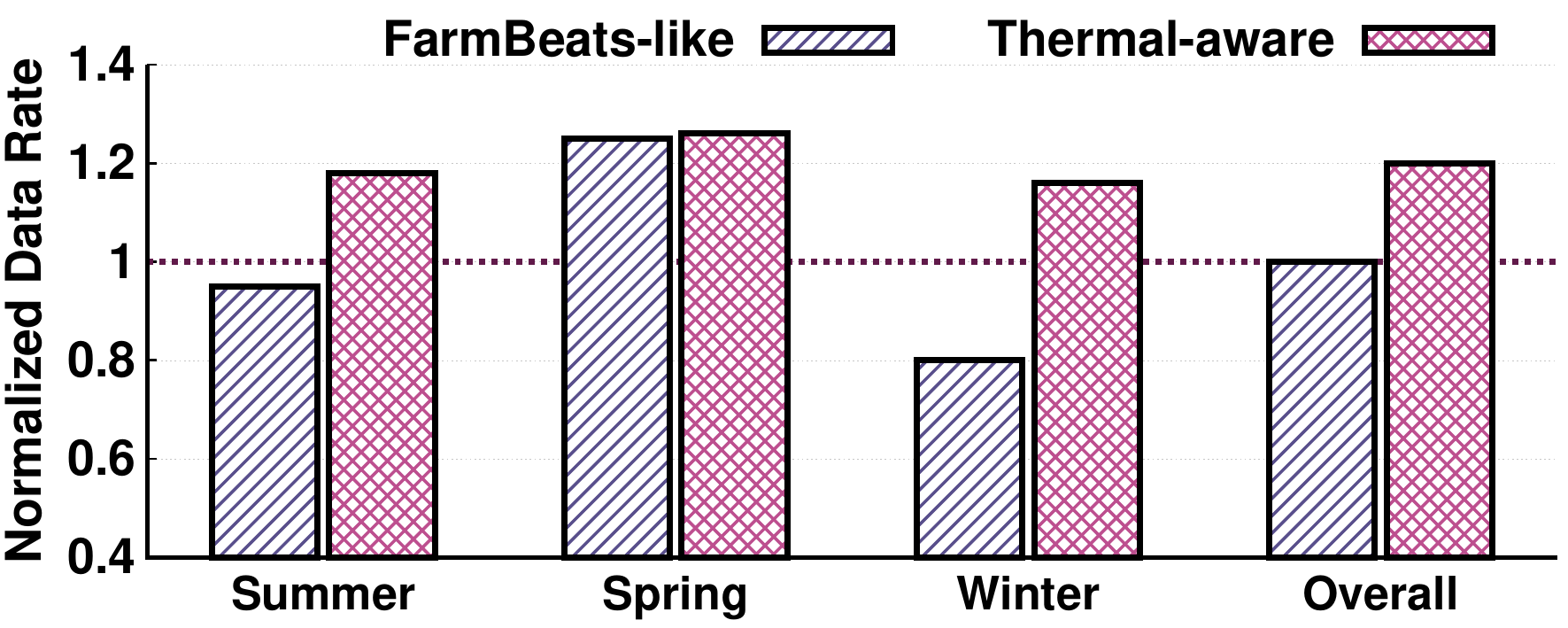}
    \vspace{-0.35cm}
    \caption{\emph{Performance comparison against Farmbeats. A thermal-aware operation outperforms Farmbeats thermal-agnostic design during the summer and winter months. Both have the same performance during spring season.}}
    \label{fig:case1_agri}
    \vspace{-0.6cm}
\end{figure}

\vspace{-0.1cm}
\subsection{Sensor Data Acquisition}
\vspace{-0.05cm}
We evaluate the performance of our \emph{thermal-aware} approach against Farmbeats~\cite{farmbeats}. 
Farmbeats attempts to minimize the data gaps by varying the duty cycle of the data acquisition.
\citet{farmbeats} detail the hardware specifications of the system.
The system is powered by two solar panels of 60W each. Solar panels are connected to four 12V-44Ah batteries connected in parallel. 
The processing component is a Raspberry Pi 4B that consumes 2.7W in idle and roughly 7W at maximum. 
The environmental sensors are interfaced with the base station through a 802.11b router, that consumes 20W power at maximum with a base power of 3W.
We assume a linear relationship between the router power and sensor data acquisition rate for the purpose of this case study.
However, as shown in the FarmBeats paper (Fig 6a), there is no enclosure for batteries. In our case study, we design an enclosure for the system that we use for both the FarmBeats-like and our proposed thermal-aware approach. This case study essentially compares the performance of thermal-agnostic FarmBeats-like system and a thermal-aware operation for data acquisition. In the former case, we only change the rate based on the available energy. In the later case, we leverage scheduling-energy effect to get better performance. 

Figure~\ref{fig:case1_agri} shows the data rate achieved under \emph{thermal-aware} and thermal-agnostic Farmbeats operation. 
Both approches perform similarly during spring when the temperatures are moderate and there are no significant variations in temperature over the course of a single day. 
However, the \emph{thermal-aware} operation of the base station outperforms Farmbeats during summer and winter. 
The difference is significant during winter when thermal effects are significant. 
Overall, our \emph{thermal-aware} approach achieves 24\% higher data rate than Farmbeats'  thermal-agnostic design. 

\begin{figure}[t]
    \centering
    \begin{tabular}{cc}
    \includegraphics[width=0.51\linewidth]{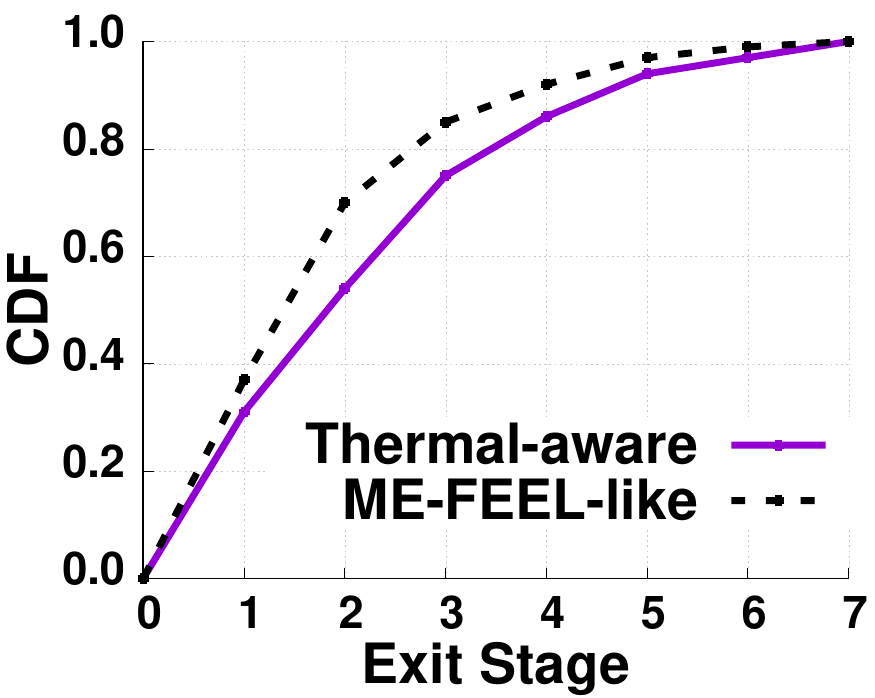} &
    \includegraphics[width=0.30\linewidth]{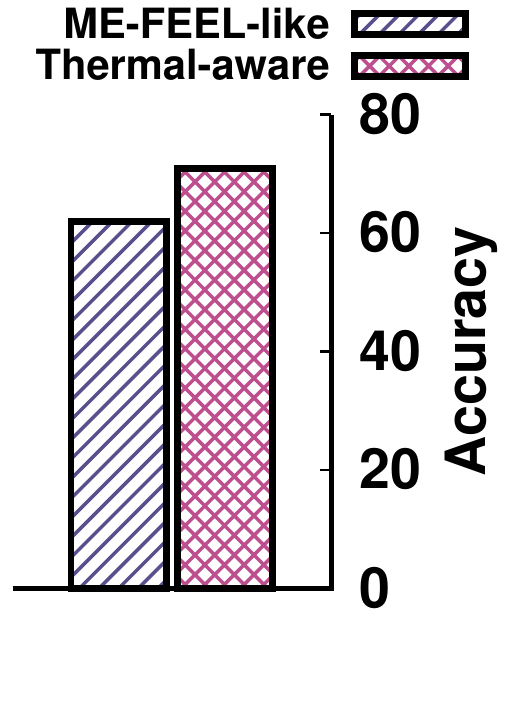} \\
    (a) Work Rate&
    (c) Accuracy\\
    \end{tabular}
     \vspace{-0.35cm}
    \caption{\emph{Performance comparison against a modified MultiExit Federated Edge Learning (ME-FEEL) approach~\cite{edge-federated2}. A thermal-aware operation outperforms ME-FEEL by exiting in a higher stage leading to a higher training accuracy.}}
    \label{fig:case2_perform}
    \vspace{-0.65cm}
\end{figure}

\subsection{Federated Learning at the Edge}
There is significant prior work on leveraging federated learning on resource-constrained edge devices~\cite{edge_federated, edge-federated2, guo2022bofl}.
These approaches aim to maximize the accuracy of the trained model under energy availability constraints at each device and constraints on the bandwidth available to upload the local model to the parameter server. 
However, these approaches do not explicitly consider how these constrained resources might be further impacted when exposed to a wide range of extreme environments.
This case study shows that by jointly managing electrical and thermal energy, we can achieve better performance, both in terms of energy efficiency and work rate for such environmentally-powered computer systems. 

\noindent \emph{{\bf Application Setup.}} Our baseline for this case study is the Multi-Exit Federated Edge Learning (ME-FEEL) approach proposed by Tang et al.~\cite{edge_federated}, which uses a modified ResNet18~\cite{ResNet-18} deep learning model that enables exiting at any of its seven layers. 
\citet{edge_federated} profile the training time for each exit stage that we leverage in our simulation.
Our application scenario consists of periodic rounds where the edge device trains its model and uploads the results to the server within a certain time threshold. 
The edge device picks an exit stage based on the energy availability at the device. 
A higher exit stage indicates more work done and vice versa. Since the amount of energy in both cases is the same, a higher work-rate means higher energy efficiency. Furthermore, since the system selects the exit stage based on the available energy, it is available for 100\% of the rounds.
We evaluate two scenarios where an application maximizes its work rate and energy-efficiency over a long period. 

\noindent\emph{{\bf Optimizing Work Rate.}} 
As illustrated in~\cite{edge_federated}, a higher exit stage yields higher accuracy and application aims to exit at the highest stage possible. 
To achieve this goal, the application must prioritize the work rate over system availability and energy-efficiency. 
Figure~\ref{fig:case2_perform} compares the performance of the ME-FEEL's thermal-agnostic approach to our \emph{thermal-aware} approach.  
A \emph{thermal-aware} approach results in a higher probability of exiting at a higher stage. 
As a result, a \emph{thermal-aware} model training has a higher accuracy. 
However, in attempting a higher exit stage and increased work rate, the node may run out of energy and not be able to participate in some rounds. 
Thus, despite high accuracy, the model may not be trained on the latest data. 
Our approach achieves an accuracy of 71\% versus 62\% for the baseline, representing a 14.5\% improvement.

\section{Related Work}
\label{sec:related}

\noindent
\textbf{Energy-harvesting Sensor Systems.} 
There is prior work on designing environmentally-powered systems that dynamically adapt their energy usage to enable perpetual operation, mostly for small-scale energy-harvesting sensor systems~\cite{perpetual,ganesan,cloudy}. 
Most of the prior work assumes ideal operating conditions, e.g., 20-25$^\circ$C and ignores thermal effects. The closest work to ours is \cite{battery-discharge-park-2005}, as it considers the effect of ambient temperature and discharge current on battery's energy-efficiency. However, it ignores other effects in our work, e.g., insulation-fan effect and scheduling-energy effect.

\noindent
\textbf{Edge AI.}
There is prior work on Edge AI that maximizes the energy-efficiency of the edge computing platforms by optimizing various power management techniques~\cite{guo2022bofl}.
However, this body of work is orthogonal and can be used in conjunction with our approach. 

\noindent
\textbf{Sustainable Clouds.} 
There is recent work on designing sustainable clouds powered by renewable energy with battery storage~\cite{,greencassandra,parasol,greenslot,sharma:asplos11,hotnets21}, which focuses on adapting the workload to match variations in the energy supply but ignores thermal effects. 

\noindent
\textbf{Data Center Cooling.}
Prior work on managing heat in large-scale data centers mainly focuses on heat movement within facilities and avoiding hotspots. The work on free cooling data centers discusses the impact of ambient temperature on energy-efficiency and workload adaptation to optimize cooling efficiency~\cite{free-cooling1,free-cooling2}. However, data centers do not have large batteries and thermal effects in them are limited~\cite{thermal1}. Also, data centers focus on cooling; the heat produced is a waste that must be dissipated.  In contrast, environmentally-powered computer systems can leverage this heat in cold weather.

\noindent
\textbf{Energy Modeling.} There is work on modeling thermal energy in buildings, such as OpenStudio~\cite{open-studio}. 
However, it generally does not model battery components and its thermal effects. 

\section{Conclusion}
\label{sec:conclusion}
In this paper, we considered environmentally-powered computing system and showed that they must consider not only their electrical energy but also the thermal energy for effective  design and operation. Our evaluation showed that a season-specific design can achieve up to 35\% higher energy-efficiency than a non-optimal design while also outperforming the non-optimal design by achieving 20\% higher availability. Finally, our case studies showed that the thermal-aware operation of the systems yield an improvement of 24\% in data acquisition rate for precision agriculture application, 14\% increase in model accuracy for the federated learning at the edge, and 41\% increase in the data used for training at the edge. 

\section*{Acknowledgment}
We thank the anonymous e-Energy reviewers for their insightful comments and feedback. This research is supported by NSF grants 2230143, 2107133, 2105494, 2213636, 2211302, 1908536, 1925464, 2211888, US Army contract W911NF-17-2-0196, and CX-027429.

\balance

\end{document}